\documentclass[aps,pre,twocolumn,superscriptaddress,showpacs]{revtex4-1}
\usepackage{natbib}
\usepackage[english]{babel}
\usepackage[dvips]{graphics}
\usepackage{graphicx,epsfig}
\usepackage{amsmath}
\usepackage{color}
\usepackage{multirow}
\usepackage[normalem]{ulem}
\usepackage{booktabs}
\newcommand{\E}{\mathcal{E}}
\newcommand{\R}{\mathcal{R}}
\newcommand{\glob}{\textnormal{}}
\newcommand{\loc}{\textnormal{loc}}
\newcommand{\PR}{R}
\newcommand{\pr}{r}
\newcommand{\scc}{\textnormal{sc}}
\newcommand{\W}{\textnormal{W}}

\newcommand{\with}{\textnormal{with}}

\begin{document}
\title{Using basis sets of scar functions}
\author{F. Revuelta}
\email[E--mail address: ]{fabio.revuelta@upm.es}
\affiliation{Grupo de Sistemas Complejos
and Departamento de F\'{\i}sica,
Escuela T\'ecnica Superior de Ingenieros
Agr\'onomos, Universidad Polit\'ecnica de Madrid,
28040 Madrid, Spain}
\author{R. M. Benito}
\email[E--mail address: ]{rosamaria.benito@upm.es}
\affiliation{Grupo de Sistemas Complejos
and Departamento de F\'{\i}sica,
Escuela T\'ecnica Superior de Ingenieros
Agr\'onomos, Universidad Polit\'ecnica de Madrid,
28040 Madrid, Spain}
\author{F. Borondo}
\email[E--mail address: ]{f.borondo@uam.es}
\affiliation{Departamento de Qu\'{\i}mica,
and Instituto de Matem\'aticas (ICMAT),
Universidad Aut\'onoma de Madrid,
Cantoblanco, 28049--Madrid, Spain}
\author{E. Vergini}
\email[E--mail address: ]{vergini@tandar.cnea.gov.ar}
\affiliation{Grupo de Sistemas Complejos
and Departamento de F\'{\i}sica,
Escuela T\'ecnica Superior de Ingenieros
Agr\'onomos, Universidad Polit\'ecnica de Madrid,
28040 Madrid, Spain}
\affiliation{
Departamento de F\'isica, Comisi\'on Nacional de Energ\'ia At\'omica,
Av.~del Libertador 8250, 1429 Buenos Aires, Argentina}
\date{\today}
%
%
\begin{abstract}
We present a method to efficiently compute the eigenfunctions of
classically chaotic systems.
The key point is the definition of a modified Gram-Schmidt procedure
which selects the most suitable elements from a basis set of scar
functions localized along the shortest periodic orbits of the system.
In this way, one benefits from the semiclassical dynamical properties
of such functions.
The performance of the method is assessed by presenting an application
to a quartic two dimensional oscillator whose classical dynamics
are highly chaotic.
We have been able to compute the eigenfunctions of the system using a
small basis set. 
An estimate of the basis size is obtained from the mean participation ratio.
A thorough analysis of the results using different indicators,
such as eigenstate reconstruction in the local representation,
scar intensities, participation ratios, and error bounds,
is also presented.
\end{abstract}
\maketitle
%
%
\section{Introduction}
  \label{sec:intro}

The vast majority of methods to obtain quantum stationary
states rely on the expansion of the corresponding wave functions
in a basis set of suitable basis functions that can be made
(approximately) complete,
on which the Hamiltonian of the system is diagonalized.
The choice of the basis set is then critical for the efficiency
of the method.
This issue is particularly important in the case of heavy particle
dynamics or in the semiclassical limit, where these functions
oscillate considerably.
The situation is even worse for very chaotic or ergodic systems,
as those in which we are interested in this paper.

Several methods have been proposed in the literature.
The simplest procedure uses products of harmonic oscillator eigenfunctions,
something which works well to describe a good number of  low-lying states,
but gets progressively poor as energy increases due to anharmonicities
(see, for example, Refs.~\onlinecite{pullenedmonds81_carne84_EHP89,%
Waterland}).
Going to the other extreme, other methods have been proposed
making use of the semiclassical information derived from
quantized invariant classical structures~\cite{brack97},
that render excellent results~\cite{Davis_Blanco_Heller,bogo92}.

In this paper we investigate the feasibility of using scar functions,
localized over short periodic orbits (POs), as a basis set to efficiently
compute the eigenstates of classically chaotic Hamiltonian systems.

The term ``scar'' was introduced by Heller in a seminal paper~\cite{heller84} 
to describe the dramatic enhancement
of quantum probability density that takes place along POs in some
eigenfunctions of the Bunimovich stadium billiard,
as a result of the recurrences along the scarring orbit.
The relevance of unstable POs in the quantization of classically
chaotic systems had been previously pointed out by Gutzwiller in
his celebrated trace formula (GTF)~\cite{gutz90}.
Other fundamental contributions to the theory of scars~\cite{kaplan}
were made by Bogomolny~\cite{bogo88}, who showed how this extra density
is obtained by averaging in configuration space groups of eigenfunctions
in an energy window around Bohr--Sommerfeld (BS) quantized energies
in the $\hbar \rightarrow 0$ limit.
The corresponding phase space version using Wigner functions
was investigated by Berry~\cite{berry89b}.
Other interesting aspects of scarring,
such as the role of homoclinic and heteroclinic quantized circuits~\cite{Tomsovic,us},
the influence of bifurcations (in systems with mixed dynamics)~\cite{Prado},
the scarring of individual resonance eigenstates in open systems~\cite{open},
or relativistic scarring \cite{emc2}
have also been discussed in the literature.
Scars have also been experimentally observed in many different contexts,
including microwave cavities~\cite{ scarsexp},
semiconductor nanodevices~\cite{nano},
optical microcavitities~\cite{optcav},
optical fibers~\cite{optfib}, 
and graphene sheets~\cite{graphene}.

Different methods have been described in the literature
to systematically construct functions localized on unstable POs
(hereafter called scar functions).
Polavieja \emph{et al}.~averaged groups of eigenstates using the short-time
true quantum dynamics of the system~\cite{pola94}.
Vergini and coworkers used the short POs theory~\cite{ver00}
and obtained scar functions by combination of resonances of POs
over which condition of minimum energy dispersion is imposed,
thus including the semiclassical dynamics around the scarring PO
up to the Ehrenfest time~\cite{ver01}.
Sibert \emph{et al}.~\cite{sibert06} and Revuelta \emph{et al}.~\cite{Fabio}
extended the method to smooth potential systems.
Also, Vagog \emph{et al}.~extend to unstable POs the asymptotic boundary
layer method to calculate stable microresonator localized modes~\cite{scho09}.
These scar functions appear not only well localized in configuration
and phase space, but they also present a very low dispersion in
energy~\cite{ver08}, and this property makes of them good candidates
\emph{a priori}  to form an efficient basis set for the calculation
of the eigenstates of classically chaotic systems.
An additional advantage of using this kind of basis functions,
which are based on dynamical information,
is that they should allow an easy and straightforward identification
of the underlying invariant classical structures that are relevant
for the semiclassical description of individual states of a chaotic system.

In this paper we introduce a new method to construct basis sets
formed by the scar functions described before~\cite{ver01,sibert06,Fabio}
that can be used to efficiently compute the eigenvalues and
eigenfunctions of classically chaotic systems with smooth potentials.

This methods exploits a simple semiclassical idea,
based on the well known Weyl law for closed systems,
which gives an intuitive explanation of how the quantum states
of a system ``fill'' the corresponding phase space~\cite{brack97}.
Put in numerical terms, the associated volume divided by that of a
Planck cell (that taken by a single state) gives a semiclassical
estimation of the generated Hilbert space size,
%
$$ N_\E = \frac{\int \int d\textbf{p} d\textbf{q} \;
\Theta[\E-\mathcal{H}(\textbf{p}, \textbf{q})]}
{(2 \pi \hbar)^d}, $$
where $d$ is the dimensionality of the problem,
and $\Theta$ the Heaviside function.
The application of this prescription, i.e.~the calculation of the
phase space integral in the above expression, is particularly simple
when one aims at calculating all states up to a given energy, $\E$,
which then provides an easy way to compute a minimum bound to the
dimension of the required basis set.
Obviously, it is always advisable to increase this number a little
bit to account for the border effects, in order to obtain a better
description of the states localized on this region of phase space.
This type of strategy has long been proposed in the literature.
For example, Heller \emph{et al}.~\cite{Davis_Blanco_Heller} choose
to fill the relevant phase space up to the considered energy
with coherent states or linear superpositions of such states placed
along quantized trajectories.
Notice that the usual basis sets, constructed for example with
(orthogonal) products of harmonic oscillator functions on each coordinate,
are less efficient since they are worse adapted to the relevant phase space,
unnecessarily extending into the classically forbidden regions.
Bogomolny~\cite{bogo92} used semiclassical  functions distributed
in a narrow crust around a given energy shell,
thus covering a phase space volume given by the surface of the
mean energy shell times the energy width of the functions.
Our method is similar in spirit to this one,
and the required phase space up to a given energy is then filled up
with a succession of overlying layers, one in top of each other,
in an onion-like fashion.
Very recently~\cite{Carlo}, basis sets of scar functions defined
over short POs have been used to compute the eigenstates of
the evolution operator in open quantum maps.
In addition to showing a great performance for this task, they have
proven a very powerful tool for the analysis of the behavior
of these kind of systems.
The fact that our scar functions are defined with a very low energy
dispersion makes that they fill very effectively,
i.e.~with smaller basis sizes, the phase space.
Notice that some complications arise when constructing the basis elements,
due to the overlap existing among the scar functions.
Finally, let us remark that the kind of basis sets proposed here can
be considered as optimal from a semiclassical point of view,
since they minimize the dispersion by making a time evolution until Ehrenfest time [see Eq.~(\ref{eq:ehrenfest}) below].

The performance of the method is illustrated with an application to a
highly chaotic coupled quartic oscillator with two degrees of freedom
that has been extensively studied in connection with quantum
chaos~\cite{pullenedmonds81_carne84_EHP89,Waterland,bohi93}.
We show how our method is able to accurately compute the~$\sim$2400
low-lying eigenfunctions of the quartic oscillator using
a basis set consisting only of $\sim$2500 elements constructed
over 18 POs of the system.
Furthermore, we demonstrate that the method can be used
to compute the eigenfunctons of the system in a small energy window 
using a basis set whose size is of the same order of magnitude
as the number of computed eigenfunctions.
An estimate of this basis size is obtained from the mean participation ratio, 
and it results much smaller than those used in other methods.
The extension to systems of higher dimensionality is straightforward.

The organization of the paper is as follows.
In Sect.~\ref{sec:system} we introduce the system that has been
chosen to study.
Section~\ref{sec:method} is devoted to the description of the method
that we have developed, which is based on two central or key pillars.
First, we use a general procedure able to construct localized
functions along unstable POs for Hamiltonian systems with
smooth potentials.
Second, we use a \emph{selective} Gram-Schmidt method (SGSM) to
select a linearly independent scar function subset
from an overcomplete set within a given energy window,
thus obtaining by direct diagonalization of the corresponding Hamiltonian
matrix the desired eigenenergies and eigenfunctions with a
great degree of accuracy using standard routines.
We also describe in this section the different mathematical tools that
will be later used in the analysis of the quality of our results.
They are: 1) local representation functions,
2) scar intensities or contribution of each PO to the emergence of an
individual eigenfunction, and
3) participation ratios, from which an approximated idea of the number of
basis elements needed to reconstruct an eigenfunction can be obtained.
In Sect.~\ref{sec:results} we present and discuss the results obtained
in the calculation of the eigenstates of the system.
Finally, in Sect.~\ref{sec:summary} we summarize the main
conclusions of our work and make some final remarks.

\section{System}
\label{sec:system}

Our system consist of a particle of unit mass moving in  a quartic potential
on the $x-y$ plane
%
\begin{equation}
\mathcal{H}(P_x,P_y,x,y)=\frac{P_x^2+P_y^2}{2}+\frac{x^2 y^2}{2} +
\frac{\beta}{4}(x^4+y^4),
\label{eq:1}
\end{equation}
with the parameter $\beta=1/100$.
This Hamiltonian has been very often used in studies concerning quantum chaos~\cite{pullenedmonds81_carne84_EHP89,Waterland,bohi93,ver09}.
%
\begin{figure}
\includegraphics{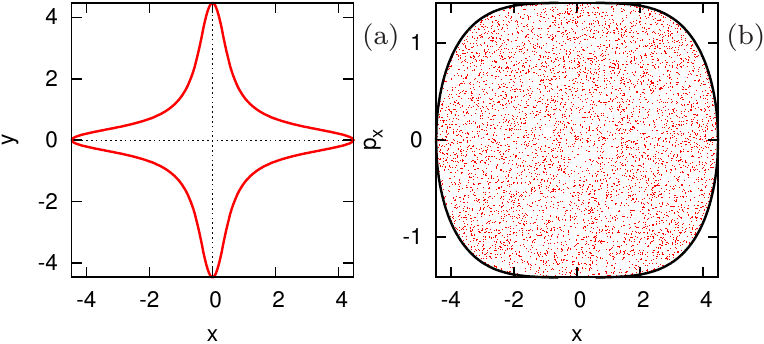}
\caption{(color online)
(a) Equipotential line for the quartic oscillator corresponding
to $\E=1$. \\
(b) $y=0$ with $P_y>0$ Poincar\'e surface of section for a typical
trajectory of the system at $\E=1$.
No signs of regular motion are apparent.}
\label{fig:1}
\end{figure}
A plot of the equipotential line $\E=1$ is shown in Fig.~\ref{fig:1}(a).
The corresponding dynamics are highly chaotic;
notice that no signs of invariant tori can be identified in Fig.~\ref{fig:1}(b),
where the $\{ y=0, P_y>0 \}$ Poincar\'e surface of section (SOS) for a typical
trajectory at $\E=1$ is shown.
However, some stable POs exist~\cite{dahlq90,simo11}, although the area
of their stability regions are negligible for all practical purposes.
%
\begin{figure}
\includegraphics{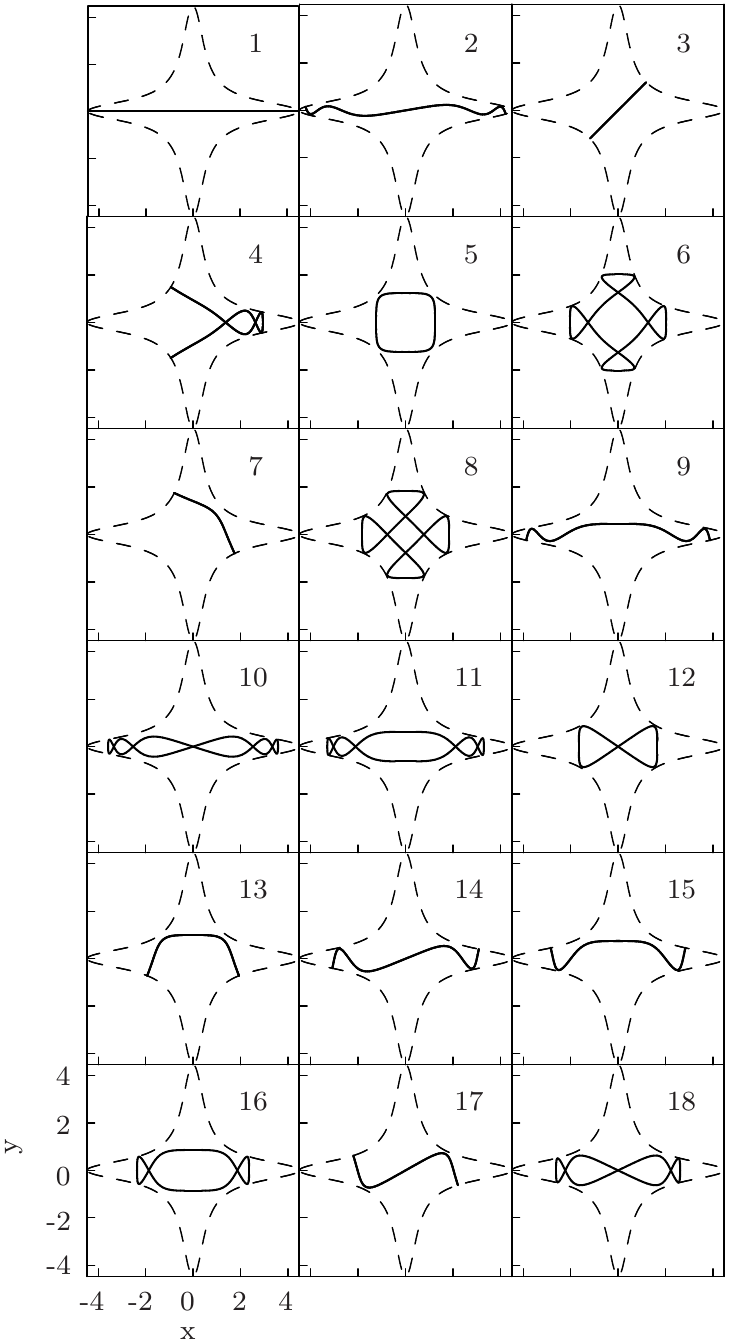}
\caption{Unstable periodic orbits for the quartic oscillator~(\ref{eq:1})
at energy $\E=1$ relevant for this work.
The corresponding equipotential line has also been drawn in dashed line.}
\label{fig:2}
\end{figure}
Another interesting property of Hamiltonian (\ref{eq:1}) is that,
due to the fact that the potential is homogeneous, it is mechanically similar.
This implies that any trajectory, $(x_t,y_t,P_{x,t},P_{y,t})$,
at a given energy, $\E$,
can be scaled to another, $(x'_{t'},y'_{t'},P'_{x,t'},P'_{y,t'})$, at a different energy, $\E'$,
by using the simple scaling relations
%
\begin{equation}
\begin{array}{ccc}
x'_{t'} & = & \displaystyle \left(\frac{\E'}{\E}\right)^{1/4} x_t, \quad
\displaystyle P'_{x,t'} = \left(\frac{\E'}{\E}\right)^{1/2} P_{x,t},\\
y'_{t'} & = & \displaystyle \left(\frac{\E'}{\E}\right)^{1/4} y_t, \quad
\displaystyle P'_{y,t'} = \left(\frac{\E'}{\E}\right)^{1/2} P_{y,t},
\label{eq:2}
\end{array}
\end{equation}
where $t'=(\E'/\E)^{-1/4} t$.
Notice, that the above relations imply that the structure of the phase space is
the same for all values of the energy.
Furthermore, from these expressions it can be easily shown that the classical action,
$S_t=\int_0^t d\tau \, [P_{x,\tau}^2+P_{y,\tau}^2]$, scales as
%
\begin{equation}
S'_{t'}=\left(\frac{\E'}{\E}\right)^{3/4} S_t
\label{eq:3}.
\end{equation}
Notice that the period of a PO fulfills \linebreak
$T'=(3 /4) S_{T}/\E$, since $T=dS/dE$.

Finally, we present in Fig.~\ref{fig:2} the eighteen POs for the system
described by Hamiltonian (\ref{eq:1}) that will be used in the
calculations of this paper.
They have been chosen as the semiclassically most relevant ones,
in the sense that they are short, symmetric and not too unstable
(see discussion below, in Sect.~\ref{sec:BS}).
%
\section{Method}
\label{sec:method}

In this Section we describe the method that is used in our calculations.
This description is made in three steps.
First, we define the scar functions that are used to construct our basis sets.
Second, we describe the procedure by which the elements of the basis set
are selected and computed.
And third, we introduce the mathematical tools that will be used
to analyze the characteristics and quality of our results.
%
\subsection{The scar functions}
\label{sec:scar}

This subsection consists of four parts.
We first introduce (auxiliary) tube functions.
Then some attention is paid to the details of the BS quantization procedure
on the PO.
In the third part, we discuss the actual calculations of the scar functions,
which are obtained by dynamically improving the tube functions,
and constitute the primary ingredient of our basis elements.
Finally, some examples of scar functions are presented and discussed.
%
\subsubsection{The tube functions} \label{sec:tube}

Some auxiliary tube functions are first defined as
%
\begin{equation}
\psi_n^{\rm tube}(x, y) =
\int_0^T \! dt \, e^{i \E_n t/\hbar} \, \phi (x,y,t),
\label{eq:4}
\end{equation}
where $T$ is the period of the PO, and $\E_n$ is the associated
BS quantized energy (see next subsection).
The wave function $\phi(x,y,t)$ is a suitable wave packet,
whose dynamics is forced to stay on the neighborhood of the scarring PO,
$(x_t,y_t,P_{x,t},P_{y,t})$.
For this purpose, we use a frozen Gaussian~\cite{Heller4,Littlejohn}
centered on the trajectory, that can be expressed as
%
\begin{eqnarray}
\phi (x,y,t)=
\exp \left\{ -\alpha_{x} (x-x_t)^2 -\alpha_{y} (y-y_t)^2 + \right.
\nonumber \\
\left. (i/\hbar) [P_{x,t} (x-x_t) + P_{y,t} (y-y_t)]+ i \gamma_t \right\},
\label{eq:5}
\end{eqnarray}
where $\alpha_{x}$ and $\alpha_{y}$ define the widths along the two axis.
In our calculation we take, for simplicity, $\alpha_{x}=\alpha_{y}=1$,
which is an adequate value for the problem that we are considering here.
The phase
%
\begin{eqnarray}
\gamma_t & = & \frac{1}{\hbar}\int_0^t d\tau \, \left(P^2_{x,\tau} + P^2_{y,\tau} \right) \;
- \frac{\pi}{2} \mu_t \nonumber \\
& = & \frac{S_t}{\hbar} - \frac{\pi}{2} \mu_t
\label{eq:6}
\end{eqnarray}
is the difference between the dynamical phase of the orbit [cf.~Eq.~(\ref{eq:3})]
and a topological phase, proportional to the function $\mu_t$,
which can be calculated by applying the Floquet's theorem~\cite{yak75}.
For this purpose, the transversal motion is decomposed in the vicinity of a PO
in two terms:
one purely hyperbolic, describing the dilatation-contraction that takes place
along the directions of the associated invariant manifolds,
and another one, periodic in time, which is described by the matrix $F(t)$
introduced in Eq.~(11) of Ref.~\onlinecite{ver01}.
After one period of time, $\mu_T \equiv \mu$ is given by the winding number of the PO,
which equals the number of half turns made by the manifold directions
as they move along the PO.
Also, this number is equal to the number of self-conjugated points plus the number
of turning points existing on the PO.

In general, $\mu_t$ can be calculated as follows. First, the
transversal monodromy matrix of the PO is computed and
diagonalized. Recall that this matrix describes the linearized
motion in the vicinity of a PO, and its eigenvectors, $\xi_s$ and
$\xi_u$, give the directions of the stable and unstable invariant
manifolds in that region. A trajectory starting on the
stable/unstable manifold then approximates/separates from the PO
during the time evolution. The evolution of the eigenvectors
$\xi_s(0)$ and $\xi_u (0)$ can be written, in the linear
approximation, as follows
$$ \xi_s(t)=e^{-\lambda t} F(t) \xi_s(0),\qquad
\xi_u(t)=e^{\lambda t} F(t) \xi_u(0), $$
where $\lambda$ is the stability index of the PO. After one period
of time, $\xi_s(T)$ and $\xi_u(T)$ are $e^{\lambda T}$ times
shorter and larger, respectively, than the original vectors
$\xi_s(0)$ and $\xi_u(0)$, being the factor $e^{\lambda T}$ equal
to the absolute value of the largest monodromy matrix eigenvalue.
Moreover, $\xi_s(T)$ and $\xi_u(T)$ are either parallel or
antiparallel to $\xi_s(0)$ and $\xi_u(0)$, depending on whether
the value of $\mu$ is either even or odd, respectively, i.e.~ the
eigenvalues of the monodromy matrix are positive or negative.
If they are positive (negative), the motion in the neighborhood
of the PO is hyperbolic (hyperbolic with reflection) and $F(T)$
is equal to the unit matrix, $I$ ($-I$).
Once the eigenvectors, $\xi_s(t)$ and $\xi_u(t)$, have been calculated,
we follow the evolution of the new vectors
$$\tilde{\xi}_s(t)=e^{\lambda t} \xi_s(t),\qquad
\tilde{\xi}_u(t)=e^{-\lambda t} \xi_u(t),$$
which describe an unconventional motion in the vicinity of an
unstable PO without hyperbolicity.
For instance, let $z_{\rm PO}(t)$ be the points of the PO
as a function of time.
Then, the evolution of a neighbor point
$z(0)=z_{\rm PO}(0)+c_s \tilde{\xi}_s(0)+c_u \tilde{\xi}_u(0)$,
with $| c_s |, \; | c_u | \ll 1$, is given by
$z(t)=z_{\rm PO}(t)+c_s \tilde{\xi}_s(t)+c_u \tilde{\xi}_u(t)$.
Finally, $\mu_t$ can be obtained by
following the angle swept by any of the previous vectors. Let us
remark, that the value of $\mu_t$ is not canonically invariant, in
contrast with $\mu$.
%
\subsubsection{The Bohr-Sommerfeld quantization rules}
\label{sec:BS}

The smoothing process implicit in the integration on Eq.~(\ref{eq:4}) renders
a function with the probability density well localized along the PO.
This localization effect over the PO is maximized, by constructive interference,
when $\gamma_t$ returns to the initial point with an accumulated phase
that is a multiple of $2\pi$.
This happens when the phase fulfills the so called BS quantization rule
%
\begin{equation}
\frac{S_T}{\hbar} - \frac{\pi}{2} \mu = 2 \pi n, \quad \quad n=0,1,2,...
\label{eq:7}
\end{equation}

At this point it is necessary to discuss the symmetry of the computed
wave functions.
The quantum eigenstates of the quartic oscillator~(\ref{eq:1}) are
classified according to the $C_{4v}$ symmetry group,
which has five irreducible representations (IRs),
four of which ($A_1, B_1, A_2, B_2$) are one dimensional,
and the other one ($E$) two dimensional.
An elegant way to deal with this problem is to refer everything
to the fundamental domain defining the potential.
In our case this domain consists of the 1/8 region bounded
by one semiaxis and the neighbor semidiagonal in the case of the
one dimensional representations, $A,B$,
and the 1/4 region between the two semiaxis in the case of the
two dimensional one, $E$.
To translate this into semiclassical arguments, the POs must be
desymmetrized by ``folding'' the original trajectories into
the fundamental domain.
In this way, eigenfunction symmetry characteristics turn into
boundary conditions,
Dirichlet ($\psi=0$) or Neumann ($\partial_\perp \psi=0$),
at the axis ($x,y=0$) and the diagonals ($x=\pm y$).
When dealing with POs, this is equivalent to introducing
``artificial'' hard walls boundaries in both the axis and the diagonals,
which has two effects.
First, they reduce the length, and then the topological
(without the contributions arising from the desymmetrization)
and mechanical actions in Eq.~(\ref{eq:6}) in an integer factor of $p$,
given by the ratio between the periods of the full and desymmetrized POs.
Second, they have an additional more complicated effect in the Maslov index,
which is different for the Dirichlet and Neumann cases,
that has to be carefully taken into account.
Accordingly, the quantization condition (\ref{eq:7}) should be modified
in order to quantize a desymmetrized PO of period $T/p$,
taking into account the appropriate boundary conditions of the PO.
The new BS quantization rule then reads
%
\begin{equation}
\frac{S_T/p}{\hbar} - [N_{\rm D}-N_{\rm N}] \frac{\pi}{2}- [\mu/p+N_r] \frac{\pi}{2}
= 2 \pi n 
\label{eq:8}
\end{equation}
where $S_T/p$, $\mu/p+N_r$ and $N_r$ are, respectively, the
action, winding number, and number of reflections of the
desymmetrized PO.
Also, the number of excitations, $n$, has been 'reduced' to the
fundamental domain.
The number of Dirichlet and Neumann conditions on the wave functions
at symmetry lines (axis and diagonals) are given by $N_{\rm D}$
and $N_{\rm N}$, respectively.
Obviously, $N_r = N_{\rm D} + N_{\rm N}$.
Furthermore, it can be shown that $\mu+2 p N_{\rm D}$ equals the
Maslov index appearing in GTF~\cite{creagh90, rob91, ver00}.
A full discussion on the derivation
of Eq.~(\ref{eq:8}) can be found in Ref.~\onlinecite{ver01}.
Finally, the semiclassically allowed BS quantized energies can be
obtained by transforming Eq.~(\ref{eq:8}) with the aid of the
scaling relation (\ref{eq:3}), thus rendering
%
\begin{equation}
\E_n^{3/4}=\frac{2\pi\hbar}{(S/p)}\left [ n + \frac{\mu/p}{4} + \frac{N_{\rm D}}{2}\right],
\label{eq:9}
\end{equation}
where $S \equiv S_T$ is the action of the complete PO at energy~$\E=1$.

In Table~\ref{Table:I} we summarize all the relevant dynamical information for
the POs of Fig.~\ref{fig:2} at $\E=1$ [recall that they can be transformed to
any other value of the energy by using the scaling relations in
Eqs.~(\ref{eq:2}) and (\ref{eq:3})].
In the last column of the Table, we include an adimensional parameter defined as
%
\begin{equation}
\R = \lambda T N_s N_t ,
\label{eq:10}
\end{equation}
measuring the relative relevance of each PO, in the sense that
shorter, simpler, and less unstable orbits have lower values of $\R$.
The integers~$N_s$ and~$N_t$ take into account the spatial and time-reversal
symmetries of the orbits:
$N_s$ corresponds to the number of different POs that are obtained by application
of the $C_{4v}$ symmetry operations,
while $N_t$ is equal to $1$ when the PO is time-reversal and $2$ otherwise.
Notice that the product $N_s N_t$ equals the number of repetitions of the PO
appearing in the summation of the GTF, 
i.e.~the number of similar POs that can be constructed at the same energy, 
which depends strongly on how symmetrical the PO is as well as on the IR
that is being  considered.
%
\begin{table}
\centering
\begin{tabular}{ccccccc}
\hline \hline
PO & $S$ & $\lambda$ & $\mu$ & $N_s$ & $N_t$ & $\R$ \\
\hline
%
1    & 22.1111 & 0.1014 & 16 & 2 & 1 & 3.36 \\
2    & 22.0590 & 0.0777 & 14 & 4 & 1 & 5.16 \\
3    & 8.2945 & 0.7669 & 2 & 2 & 1 & 9.54 \\
4    & 26.0610 & 0.1296 & 14 & 4 & 1 & 10.12 \\
5    & 10.4568 & 0.7120 & 4 & 1 & 2 & 11.16 \\
6    & 25.0018 & 0.3842 & 12 & 1 & 2 & 14.40 \\
7    & 9.2936 & 0.6032 & 4 & 4 & 1 & 16.80 \\
8    & 24.9083 & 0.5334 & 8 & 1 & 2 & 19.93 \\
9    & 21.7969 & 0.3197 & 14 & 4 & 1 & 20.92 \\
10  & 21.3683 & 0.3639 & 12 & 2 & 2 & 23.33 \\
11  & 20.7624 & 0.4291 & 12 & 2 & 2 & 26.72 \\
12  & 12.7134 & 0.7043 & 4 & 2 & 2 & 26.88 \\
13  & 14.2519 & 0.6469 & 6 & 4 & 1 & 27.68 \\
14  & 20.0588 & 0.4671 & 10 & 4 & 2 & 28.12 \\
15  & 19.1639 & 0.5070 & 10 & 4 & 1 & 29.16 \\
16  & 17.0268 & 0.5769 & 8 & 2 & 2 & 29.48 \\
17  & 15.8266 & 0.6237 & 6 & 4 & 1 & 29.62 \\
18  & 18.2195 & 0.5473 & 8 & 2 & 2 & 29.92 \\
\hline
\end{tabular}
\caption{Classical action $S$, stability index $\lambda$,
winding number $\mu$, spatial and time reversal numbers $N_s, N_t$,
and relevance, $\R$, measured with Eq.~(\ref{eq:10}),
for the POs of the quartic oscillator (\ref{eq:1})
shown in Fig.~\ref{fig:2} at $\E=1$. 
}
\label{Table:I}
\end{table}

\begin{figure}
\includegraphics{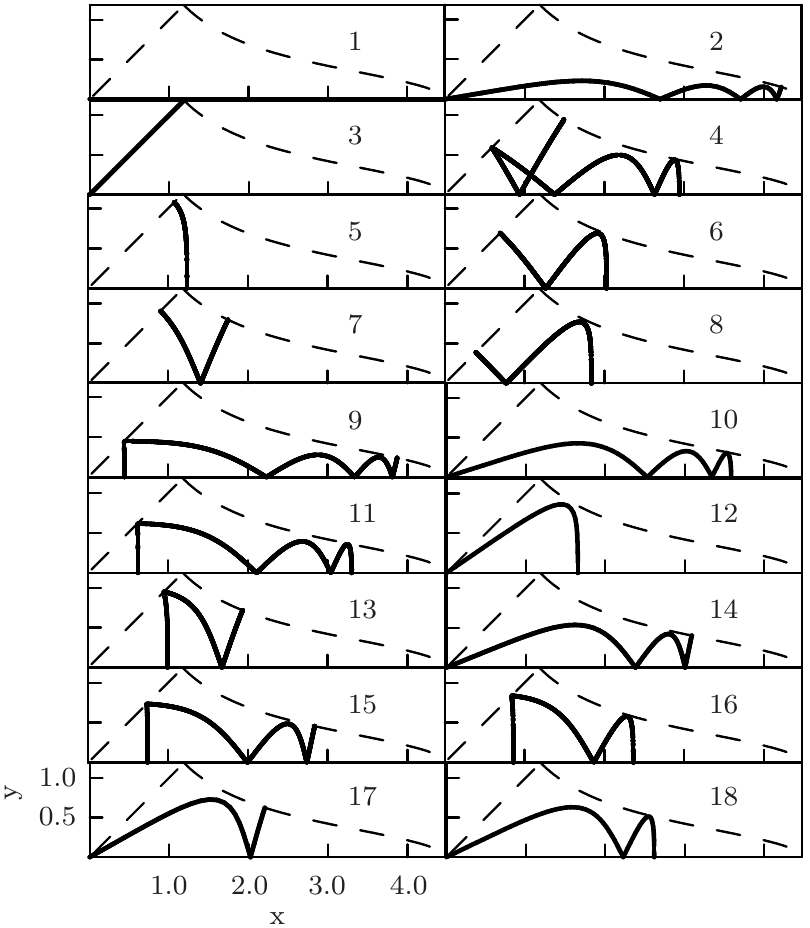}
\caption{Same as Fig.~\ref{fig:2} for the desymmetrized POs in the fundamental
domain associated to the one dimensional irreducible representations.}
\label{fig:3}
\end{figure}

\begin{figure}
\includegraphics{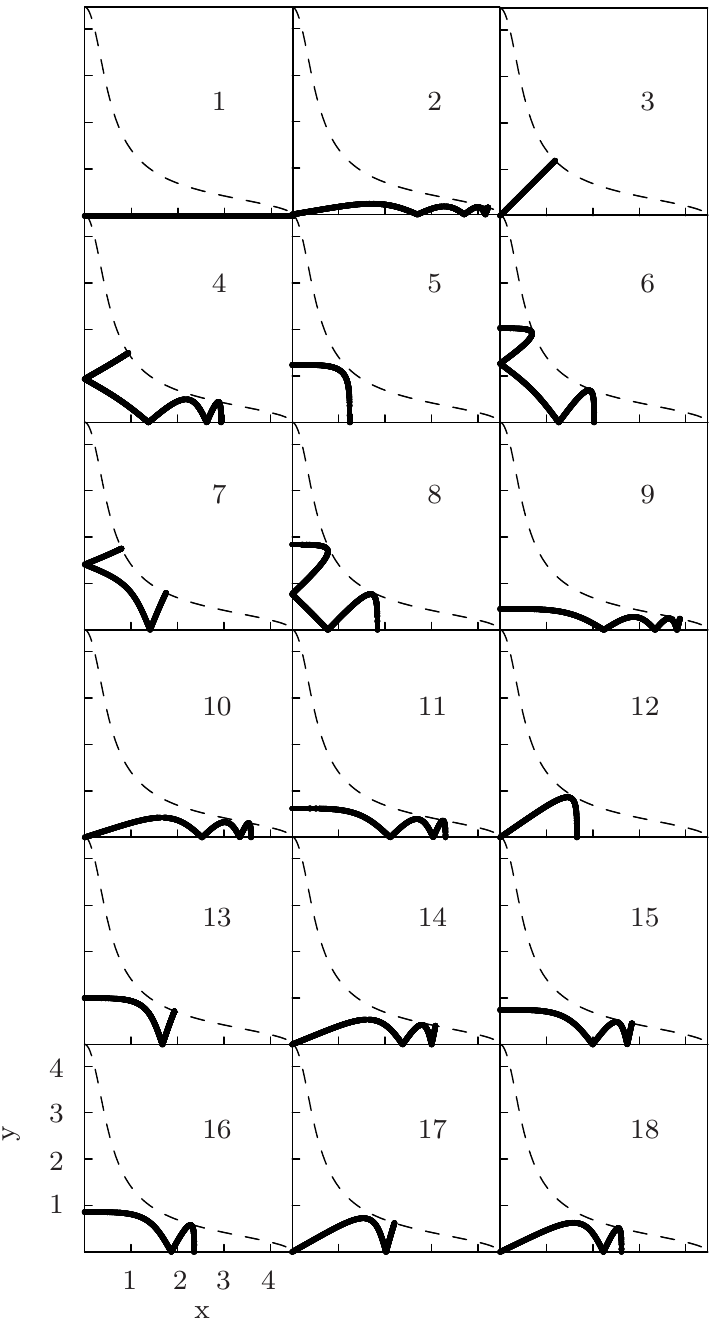}
\caption{Same as Fig.~\ref{fig:3} for the two dimensional irreducible
representation, $E$.}
\label{fig:4}
\end{figure}

Let us consider now the effect of desymmetrization.
For the one dimensional IRs, the system is desymmetrized by reducing
the configuration space to the region $x \ge y \ge 0$.
The corresponding desymmetrized POs are shown in Fig.~\ref{fig:3}.
For the two dimensional representation, the desymmetrized configuration
space corresponds to the region $x , y \ge 0$,
and the associated POs are plotted in figure Fig.~\ref{fig:4}.
The corresponding information: $p$, $N_{\rm D}$, and $N_{\rm N}$,
is given in Tables~\ref{Table:II} and \ref{Table:III}, respectively.
In Table~\ref{Table:II} the reflections at the axis $x=0$ and the diagonal $x=y$
are considered, in addition to a reflection at the $y$ axis for orbit $1$.
In Table~\ref{Table:III} the reflections at the $x$ and $y$ axis are considered.
We have separated the data in $E_1$ and $E_2$ components,
corresponding to the cases symmetric with respect to the axis $x$ or $y$
and antisymmetric with respect to the axis $y$ or $x$, respectively.
As POs numbers $2$, $3$, $10$, $12$, $14$, $17$ and $18$ arrive at the origin
$x=y=0$ forming a non-zero angle (see Figs.~\ref{fig:3} and \ref{fig:4}),
it is necessary in such situation to include reflections at the axes $x$ and $y$
simultaneously.

Once the BS energies are calculated with the aid of the desymmetrized
condition (\ref{eq:9}), the corresponding tube functions over the full PO
(given in Fig.~\ref{fig:2}) are computed using Eq.~(\ref{eq:4}).
These wave functions have $n$ nodes in the desymmetrized region of the
potential and $N_{\rm D}$ nodes along the symmetry lines,
and are real only if the PO shows time-reversal symmetry.
If this is not the case, a real function can always be obtained by combination
of the tube functions obtained with two gaussian wave packets given by Eq.~(\ref{eq:5})
running clock and counterclockwise along the orbit, respectively.
Notice, however, that the time-reversal wave functions do not belong,
in general, to any of the IRs of the system.
Again, this represents no problem since proper symmetry-class adapted tube functions
can be constructed combining the different wave functions obtained when
the elements of the $C_{4v}$ symmetry group act on the previously defined (real)
tube functions.
%
%
\begin{table}
\centering
\begin{tabular}{ccccccccccc}
\hline \hline
& & \multicolumn{2}{c}{$A_1$} & \multicolumn{2}{c}{$A_2$} & \multicolumn{2}{c}{$B_1$}
& \multicolumn{2}{c}{$B_2$} \\
\hline
PO & $p$ & $N_{\rm D}$ & $N_{\rm N}$ & $N_{\rm D}$ & $N_{\rm N}$ & $N_{\rm D}$
& $N_{\rm N}$ & $N_{\rm D}$ & $N_{\rm N}$ \\
\hline
1 & 2 & 0 & 1 & -- & -- & 0 & 1 & -- & -- \\
2 & 2 & 0 & 8 & 8 & 0 & 0 & 8 & 8 & 0 \\
3 & 2 & 0 & 2 & -- & -- & -- & -- & 2 & 0 \\
4 & 2 & 0 & 9 & 9 & 0 & 2 & 7 & 7 & 2 \\
5 & 4 & 0 & 2 & 2 & 0 & 1 & 1 & 1 & 1 \\
6 & 4 & 0 & 4 & 4 & 0 & 1 & 3 & 3 & 1 \\
7 & 2 & 0 & 3 & 3 & 0 & 1 & 2 & 2 & 1 \\
8 & 4 & 0 & 4 & 4 & 0 & 1 & 3 & 3 & 1 \\
9 & 2 & 0 & 9 & 9 & 0 & 2 & 7 & 7 & 2 \\
10 & 2 & 0 & 7 & 7 & 0 & 0 & 7 & 7 & 0 \\
11 & 2 & 0 & 8 & 8 & 0 & 2 & 6 & 6 & 2 \\
12 & 2 & 0 & 3 & 3 & 0 & 0 & 3 & 3 & 0 \\
13 & 2 & 0 & 5 & 5 & 0 & 2 & 3 & 3 & 2 \\
14 & 2 & 0 & 6 & 6 & 0 & 0 & 6 & 6 & 0 \\
15 & 2 & 0 & 7 & 7 & 0 & 2 & 5 & 5 & 2 \\
16 & 2 & 0 & 6 & 6 & 0 & 2 & 4 & 4 & 2 \\
17 & 2 & 0 & 4 & 4 & 0 & 0 & 4 & 4 & 0 \\
18 & 2 & 0 & 5 & 5 & 0 & 0 & 5 & 5 & 0 \\
\hline
\end{tabular}
\caption{Period ratio $p$, and number of Dirichlet, $N_{\rm D}$,
and Neumann, $N_{\rm N}$, boundary conditions
relevant for the Bohr-Sommerfeld quantization (\ref{eq:9})
of the desymmetrized POs in Fig.~\ref{fig:3} in
the one dimensional irreducible representations.}
\label{Table:II}
\end{table}
%
%
\begin{table}
\centering
\begin{tabular}{ccccccccccccccc}
\hline \hline
& & \multicolumn{2}{c}{$E_1$} & \multicolumn{2}{c}{$E_2$} & &
& & \multicolumn{2}{c}{$E_1$} & \multicolumn{2}{c}{$E_2$} \\
\hline
PO & $p$ & $N_{\rm D}$ & $N_{\rm N}$ & $N_{\rm D}$ & $N_{\rm N}$ & &
PO & $p$ & $N_{\rm D}$ & $N_{\rm N}$ & $N_{\rm D}$ & $N_{\rm N}$ \\
\hline
1 & 2 & 1 & 0 & -- & -- & & 10 & 2 & 1 & 6 & 6 & 1 \\
2 & 2 & 1 & 7 & 7 & 1 & & 11 & 2 & 1 & 5 & 5 & 1 \\
3 & 2 & 1 & 1 & 1 & 1 & & 12 & 2 & 1 & 2 & 2 & 1 \\
4 & 2 & 2 & 5 & 5 & 2 & & 13 & 2 & 1 & 2 & 2 & 1 \\
5 & 2 & 1 & 1 & 1 & 1 & & 14 & 2 & 1 & 5 & 5 & 1 \\
6 & 2 & 3 & 3 & 3 & 3 & & 15 & 2 & 1 & 4 & 4 & 1 \\
7 & 1 & 2 & 2 & 2 & 2 & & 16 & 2 & 1 & 3 & 3 & 1 \\
8 & 2 & 3 & 3 & 3 & 3 & & 17 & 2 & 1 & 3 & 3 & 1 \\
9 & 2 & 1 & 6 & 6 & 1 & & 18 & 2 & 1 & 4 & 4 & 1 \\
\hline
\end{tabular}
\caption{Same as Table~\ref{Table:II} for the desymmetrized POs
associated to the two dimensional irreducible representation,
$E$, shown in Fig.~\ref{fig:4}.
The data have been separated in two components, $E_1$ and $E_2$,
corresponding to the cases symmetric with respect to $x$ or $y$
and antisymmetric with respect to $y$ or $x$, respectively.}
\label{Table:III}
\end{table}
%
\subsubsection{Computation of the scar functions}
\label{sec:scarsub}

The scar functions are finally computed as an improved version of
the auxiliary tube functions introduced before that includes
dynamical information of the system up to Ehrenfest time, $T_E$.
By restricting the integration time in this way,
the geometrical support of the computed scar function is not
only restricted to the PO itself but it also includes small pieces
of the associated stable and unstable manifolds attached to the PO;
see, for example, discussion in Ref.~\onlinecite{sibert06}.

The scar functions are calculated by propagating the corresponding
tube functions (\ref{eq:4}), followed by a subsequent Fourier
transform for a finite lapse of time
%
\begin{equation}
\psi_n^{\rm scar} (x, y)= \int_{-T_E}^{T_E} dt \cos \left( \frac{\pi t}{2 T_E} \right)
e^{-i(\hat{\mathcal{H}}-\E_n)t/\hbar} \psi_n^{\rm tube} (x, y).
\label{eq:11}
\end{equation}
$T_E$ corresponds to the lapse of time that it takes to a typical Gaussian
wave packet to spread over the area, $A_{\rm tr}$,
in a characteristic Poincar\'e SOS of the desymmetrized system,
and it is related to the Lyapunov exponent, $\bar{\lambda}$, in the following way
%
\begin{equation}
T_E = \frac{1}{2 \bar{\lambda}} \ln \left(\frac{A_{\rm tr}}{\hbar} \right).
\label{eq:ehrenfest}
\end{equation}
In our case~$A_{tr} \sim 5.5278 \E^{3/4}$ and $\sim11.0555 \E^{3/4}$ 
for the one dimensional and two dimensional IRs, respectively, 
and $\bar{\lambda} \sim 0.3848 \E^{1/4}$,
which does not dependent on the IR
since it is associated to a generic chaotic trajectory of the system.

The cosine window in expression (\ref{eq:11}) is introduced in order
to minimize the energy dispersion, \linebreak
$\sigma=\langle \psi_n^{\rm scar} | (\hat{\mathcal{H}}- \E_n)^2 | \psi_n^{\rm scar} \rangle$,
of the scar functions~\cite{ver05},
that can be semiclassically approximated as~\cite{ver08}
%
\begin{equation}
  \bar{\sigma} = \frac{\pi}{2} \; \frac{\hbar \lambda(s_2+\lambda T_{\rm E})}
                 {(s_1+\lambda T_{\rm E})(s_2+\lambda T_{\rm E})+s_2^2},
 \label{eq:14}
\end{equation}
with $s_1 \approx 1.06078$ and $s_2=\pi/\sqrt 2 -s_1 \approx 1.16066$.
Notice that this magnitude depends on the IR through $T_E$, and then on $A_{\rm tr}$.
This procedure effectively reduces the energy dispersion
of the scar function with respect to the corresponding tube wave
function by a factor that is proportional to $\ln (A_{\rm tr}/\hbar)$.
\subsubsection{Some examples of scar functions}
\label{sec:examples}

Let us present now some representative examples of the scar functions that
are obtained for the POs in Fig.~\ref{fig:1}.
The value $\hbar=1$ will be used in all quantum computations.
Wavelets provide an efficient method to perform the time evolution 
appearing in Eq.~(\ref{eq:11}), with a precision of at least six decimal 
places~\cite{Sparks}.

%
\begin{figure}
\includegraphics{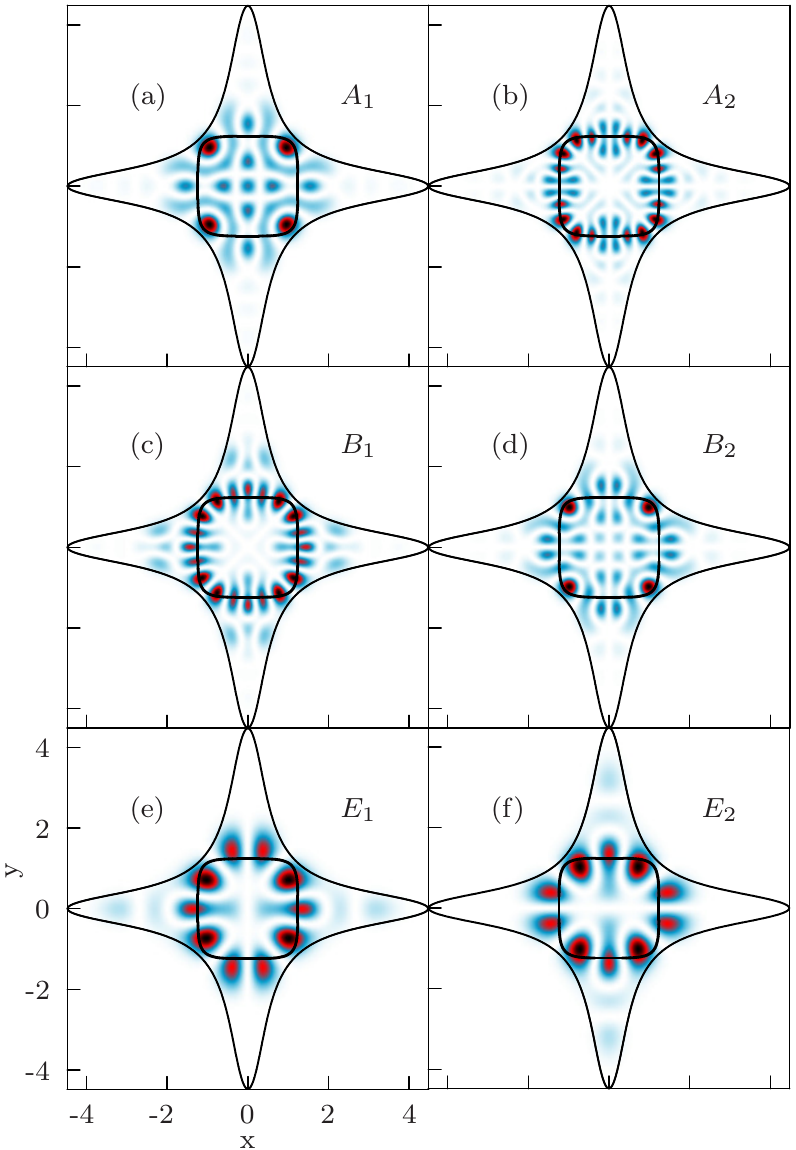}
\caption{(Color online) Probability density corresponding
to the scar functions localized over the PO number 5 of
Fig.~\ref{fig:2} with $n=2$,
for the different irreducible representations of the system.
The plot has been scaled to $\E=1$.
The scarring PO is plotted superimposed.}
\label{fig:5}
\end{figure}
In Fig.~\ref{fig:5} we show the probability density corresponding
to the scar functions constructed over the PO number 5 with an
excitation number $n=2$, for all the IRs of the system.
Since the guiding PO does not exhibit time-reversal symmetry,
time-symmetrized frozen Gaussian functions, constructed combining
wave packets~(\ref{eq:5}) running clock and counterclockwise,
have been used in the computations of Eq.~(\ref{eq:4}).
No additional spatial-symmetrization is required in order to enforce
the proper symmetry properties, since the time-symmetrized
wave functions defined in this way are invariant under the action

of the $C_{4v}$ group operations.
The BS quantized energies also necessary in the
calculations have been obtained with the aid of Eq.~(\ref{eq:9})
using the values given in Tables~\ref{Table:I}-\ref{Table:III}.
As it can be seen in the figure, the number of nodes of each of the
wave functions is different, and equal to~$p N_t (n+N_{\rm D}/2)$.
Furthermore, the functions belonging to the one dimensional IRs
are either symmetric or antisymmetric with respect to both the
axes and the diagonals, those belonging to the two dimensional IR
are on the other hand only symmetrical with respect to one axis
and antisymmetrical with respect to the other. In panel (a), the
$A_1$ scar function having $16$ nodes and local maxima over the
axis and the diagonals (Neumann boundary conditions) is shown.
On the other extreme, the $A_2$ function [panel (b)] has nodes
localized over these lines, exhibiting a total number of nodes of
$24$, $16$ of which are ``due'' to the number of excitations, $4$
come from the Dirichlet conditions at the axes, and the remaining
$4$ nodes arises from the Dirichlet conditions at the diagonals.
Similarly, the $B_1$ ($B_2$) scar functions have $20$ nodes over
the PO, $4$ of them due to the
Dirichlet conditions at the diagonals (axes).
Finally, for the $E$ symmetry we have one
Neumann condition over one axis and one Dirichlet condition over
the other, this producing a total number of nodes of $10$.

%
\begin{figure}
\includegraphics{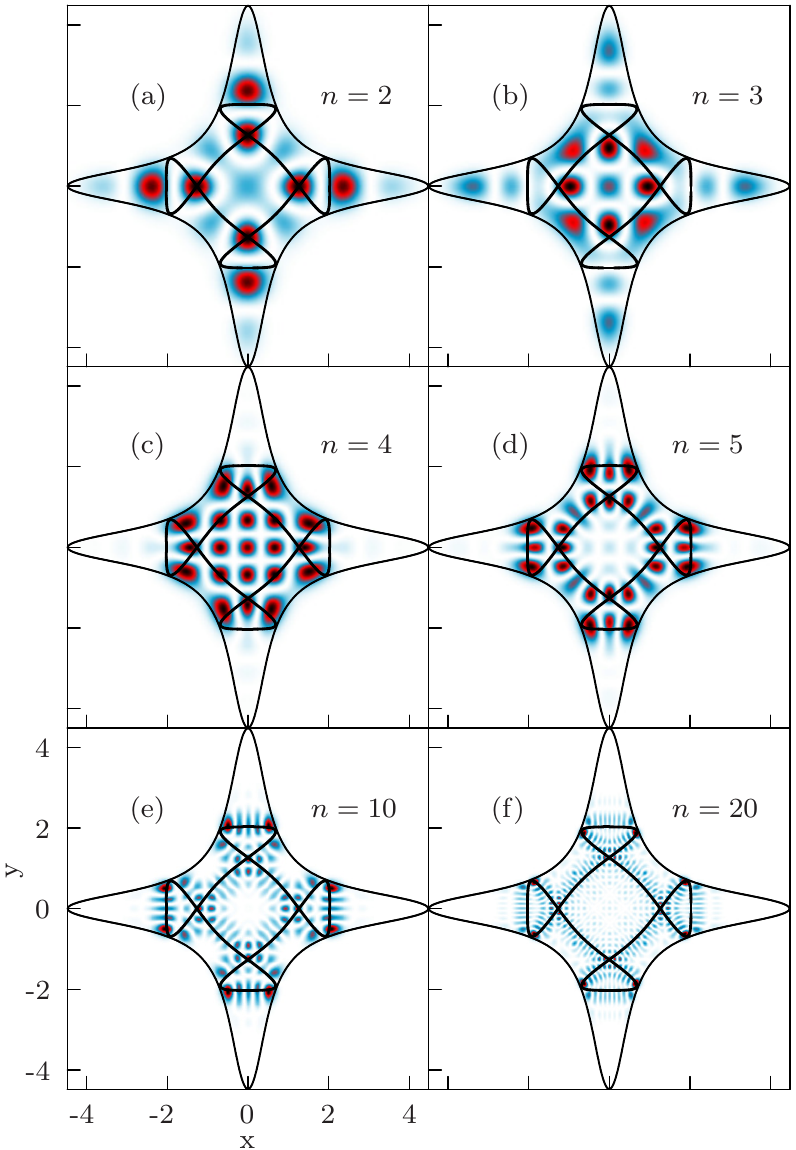}
\caption{(Color online) Scar wave functions with $A_1$ symmetry
constructed using the PO number 6 of Fig.~\ref{fig:2} for
different values of the excitation number, $n$. The scarring
PO has been superimposed.} \label{fig:6}
\end{figure}
In Fig.~\ref{fig:6} we show the $A_1$ symmetry scar functions
constructed using the PO number 6 of Fig.~\ref{fig:2} for
different values of the excitation number, $n$.
As discussed before, these functions have been constructed imposing
Neumann conditions on the boundaries of the desymmetrized POs,
so that $n$ is equal to the number of nodes in the fundamental domain,
i.e.~$8n$ in total.
The localization on the scarred PO as $n$ increases is notorious.

%
\begin{figure*}
\includegraphics[width=12cm]{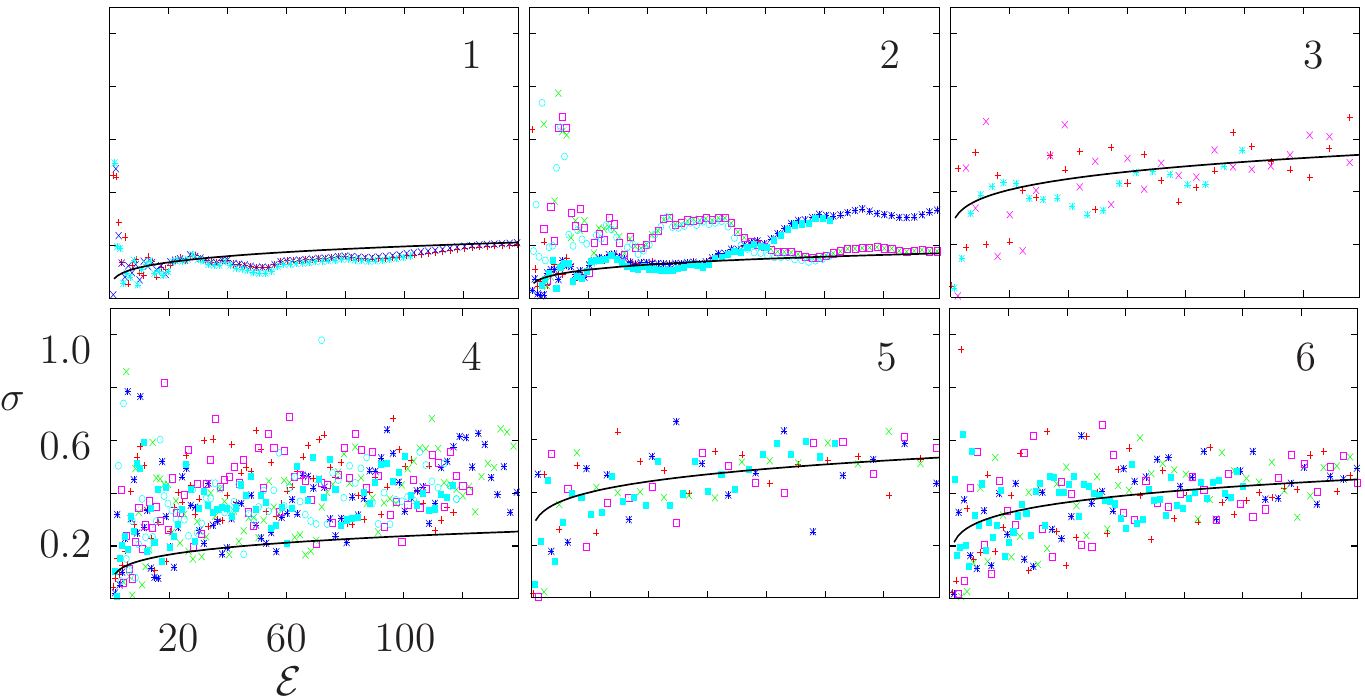}
\caption{(Color online) Dispersion of the scar functions of symmetry
$A_1$ (red plus signs), $A_2$ (green crosses), $B_1$ (dark blue stars),
$B_2$ (pink open squares), and $E$ (light blue filled squares)
constructed over the periodic orbits 1--6 shown in Fig.~\ref{fig:2}.
The results obtained with the semiclassical approximation (\ref{eq:14})
are plotted superimposed in black continuous line.}
\label{fig:sigma}
\end{figure*}
In Fig.~\ref{fig:sigma} we present some results for the energy dispersion of
our scar functions, together with the corresponding semiclassical estimation
obtained from Eq.~(\ref{eq:14}).
As can be seen, our functions are very well localized in energy,
and their dispersion grows moderately with it.
This is a key point for the aim of this paper, since it allows the definition
of a very efficient basis set.
%
\subsection{Definition of the scar functions basis set:
Selective Gram-Schmidt method}
\label{sec:GS}

The second pillar of our method is the definition of the selection
procedure of the scar functions forming the basis set,
that is subsequently used in a standard diagonalization of the
associated Hamiltonian matrix to obtain the eigenstates of the system.

To define our basis set, we have generalized the usual
Gram-Schmidt method (GSM)~\cite{lang02}, and developed a new
\emph{selective} Gram-Schmidt method (SGSM) able to choose a basis
set of linearly independent functions in a vectorial space
from a larger (overcomplete) set of
functions, that can be used to efficiently
compute chaotic eigenfunctions of our system in a given energy
window. The SGSM is specially useful when computing highly excited
eigenfunctions in a small energy window, since the size of the
basis set reduces considerably in this case.
The procedure starts from an initial set of $N$ scar functions,
$\vert \psi_j^{(0)} \rangle$, from which the SGSM selects the
minimum number of them,~$N_b \le N$, necessary to adequately
describe the Hilbert space defined by the eigenfunctions whose
energies are contained in the energy window, i.e.~the SGSM defines
a basis set in that window.
The elements of the basis set $\vert \psi_{j_i}^{(0)} \rangle$,
where subindex $i$ orders the elements according to their
semiclassical relevance (see discussion below),
are automatically selected with the aid of the conventional GSM.
Thus, associated to the basis $\vert \psi_{j_i}^{(0)} \rangle$, we
will construct an auxiliary
basis $\vert \varphi_{i} \rangle$, formed by the
orthogonalization of $\vert \psi_{j_i}^{(0)} \rangle$. For
example, if we set
$$\vert \varphi_1 \rangle = \vert \psi_{j_1}^{(0)} \rangle$$
then, the second auxiliary function $\vert \varphi_2 \rangle$ is
given by
$$\vert \varphi_2 \rangle = \frac{\vert \psi_{j_2}^{(1)}
\rangle}{\vert \psi_{j_2}^{(1)} \vert},$$
where $j_2 \ne j_1$ and
$$\vert \psi_{j_2}^{(1)} \rangle =  \vert \psi_{j_2}^{(0)} \rangle -
 \langle \varphi_1 \vert \psi_{j_2}^{(0)} \rangle \vert \psi_{j_2}^{(0)} \rangle,$$
and so on.

In our SGSM method, the selection procedure of the basis functions of a
given symmetry for the calculation of the eigenenergies, $\E$, contained
in the energy window defined by
%
\begin{equation}
\E^- < \E < \E^+,
\label{eq:15}
\end{equation}
is done automatically by using a definite set of rules,
which are based on a \emph{semiclassical selection parameter},~$\eta$.
This parameter is defined in such a way that it takes into account
in a simple form the dispersion of the scar functions,
the simplicity of the PO, and
the density of states of the system
(which is only relevant when the energy window is large).
For a given scar function,~$\eta_j$ is given by
%
\begin{equation}
  \eta_j = \rho_j [\sigma_j^2+(\delta \E_j)^2]^{1/2} T_j N_{s,j} N_{t,j},
 \label{eq:13}
\end{equation}
where $\rho_j$ is the mean density of (symmetry-class) states at the
BS energy of interest ($\E_j$), $\sigma_j$ is the dispersion of the scar function,
and $\delta \E_j$ is defined as
%
\begin{equation}
 \delta \E_j = \left\{ \begin{array}{ll}
                           \E^- - \E_j,   & {\rm if} \; \E_j < \E^- \\
                           0,                & {\rm if} \; \E^- <  \E_j <  \E^+ \\
                           \E_j  - \E^+, & {\rm if} \; \E_j > \E^+.
                           \end{array}
                  \right.
\end{equation}
Other criteria could be used to define the parameters introduced in Eqs.~(\ref{eq:10}) 
and~(\ref{eq:13}).
For example, one could drop out the function $\delta \E_j$ 
appearing in Eq.~(\ref{eq:13}) and still get quite accurate results, 
specially in large energy windows. 
However, this function is included to improve the numerical accuracy by 
reducing the boundary effects.
We thus believe that the previous equations are very straightforward in order
to substitute the contribution of the longer POs appearing in the GTF 
by the interaction of the shortest ones, following the short PO theory developed
by Vergini \emph{et al}.~\cite{ver00,ver01,ver08}, 
although other definitions are possible.

The SGSM is then defined in an algorithmic way as follows:
\begin{itemize}
\item \textbf{0.a} With the method described in
Sect.~\ref{sec:scar}, we compute all normalized scar functions,
$\vert \psi_j^{(0)}\rangle$ with the smallest values of $\R$ and
BS quantized energies $\E_j$ in the enlarged energy window
%
\begin{equation}
\E^- - 2 \bar{\sigma} < \E_j < \E^+ + 2 \bar{\sigma},
\label{eq:16}
\end{equation}
where $\bar{\sigma}$ is given by Eq.~(\ref{eq:14}) for
$\lambda \equiv \bar{\lambda}$.
This is the most time demanding step of the procedure.

It can be \emph{a priori} expected that the overlap of the
scar functions outside this enlarged window with the desired system
eigenfunctions is negligible, due to the fact that they were
constructed minimizing their energy dispersion.

\item \textbf{0.b} The semiclassical selection parameter,
$\eta_j$, of Eq.~(\ref{eq:13}), associated to each scar function
in this first approach to the final basis set, is then calculated.

\item \textbf{1.} From this initial set of scar functions,
$\vert \psi_{j}^{(0)} \rangle$, we select a smaller number of them,
$N_b \le N$, forming the basis set that is optimal for our purposes,
in the following way.
Notice that the number of scar functions calculated in this way,
$N$, should always be greater or equal to
%
\begin{equation}
\qquad \quad   N_b = N_{\rm sc}(\E^+ +2\bar{\sigma})-N_{\rm sc}(\E^- -2\bar{\sigma})
        + c_b \bar{\sigma} \rho,
\label{eq:17}
\end{equation}
where, $N_{\rm sc}(\E)$ is the semiclassical approximation to the
number of states with an energy smaller than~$\E$, and
the term $c_b \bar{\sigma} \rho$ enlarges the window size
to take into account border effects.
If this is not the case, more (longer) POs,
and consequently more scar functions, must be included at this step.

The first element of our basis set is the scar function with the
smallest~$\eta_j$ value
%
\begin{equation}
\qquad  \quad \vert \varphi_1 \rangle =
\vert \psi_{j_1}^{(0)} \rangle, \qquad \with \;
\frac{1}{\eta_{j_1}}=\max\left\{\frac{1}{\eta_j}\right\}.
\end{equation}
This choice gives priority to the scar functions which are more
localized in energy over simpler POs, i.e.~with shorter period
and being more symmetric (smaller $N_{sj}$ and $N_{tj}$).
In a similar way, we will denote from now on by $\vert \varphi_j \rangle$
the auxiliary functions necessary for the selection of the scar functions.

\item \textbf{2.a}
The remaining scar functions are then orthogonalized to function
$\vert \psi_{j_1}^{(0)} \rangle$ using the usual GSM
%
\begin{equation}
\vert \psi_j^{(1)} \rangle = \vert \psi_j^{(0)} \rangle - \langle \varphi_1^{\glob}
\vert \psi_j^{(0)} \rangle \vert \varphi_1^{\glob} \rangle, \quad j \neq j_1 .
\label{eq:19}
\end{equation}

\item \textbf{2.b} The second element of the basis set is
$\vert \psi_{j_2}^{(0)} \rangle$, where the index $j_2$
satisfies $j_2 \ne j_1$, and then
\begin{eqnarray}
\frac{\vert \psi_{j_2}^{(1)} \vert^2}{\eta_{j_2}} =
\max \left\{ \frac{\vert \psi_j^{(1)} \vert^2}{\eta_j} \right\}_{j \neq j_1},
\label{eq:21}
\end{eqnarray}
where the norm in the numerator has been introduced in order
to make the basis set elements as different as possible between them.
Indeed, notice that after the orthogonalization condition (\ref{eq:19})
the more similar $\vert \psi_j^{(0)} \rangle_{j \neq j_1}$ is to $\vert \varphi_1^{\glob} \rangle$,
the smaller the norm of the function $\vert \psi_j^{(1)} \vert_{j \neq j_1}$.
Then the auxiliary function $|\varphi_2\rangle$ is computed as
%
\begin{eqnarray}
\vert \varphi_2^{\glob} \rangle = \frac{\vert \psi_{j_2}^{(1)} \rangle}
{\vert \psi_{j_2}^{(1)} \vert}.
\label{eq:20}
\end{eqnarray}

The previous steps, 2.a and 2.b, are repeated for all the remaining basis elements
in the initial basis set of scar functions, in such a way that the $n^{th}$ step
in the procedure is defined as:

\item \textbf{n.a}
New functions are obtained by orthogonalization to the auxiliary one in the
previous step,~$\vert \varphi_{n-1}^{\glob} \rangle$,
%
\begin{eqnarray}
\vert \psi_j^{(n-1)} \rangle = \vert \psi_j^{(n-2)} \rangle -
\langle \varphi_{n-1}^{\glob}
\vert \psi_j^{(n-2)} \rangle \vert \varphi_{n-1}^{\glob} \rangle, \nonumber \\
j \neq j_1, j_2, ..., j_{n-1}. \label{eq:22}
\end{eqnarray}
\item \textbf{n.b} The $n$--th basis element, $\vert \psi_{j_n}^{(0)} \rangle$,
is then selected as the one for which
%
\begin{eqnarray}
\frac{\vert \psi_{j_n}^{(n-1)} \vert^2}{\eta_{j_n}} =
\max \left\{ \frac{\vert \psi_j^{(n-1)} \vert^2}{\eta_j} \right\}_{j \neq j_1, j_2,..., j_{n-1}},
\label{eq:24}
\end{eqnarray}
and then the next auxiliary function is constructed as
%
\begin{eqnarray}
\vert \varphi_n^{\glob} \rangle = \frac{\vert \psi_{j_n}^{(n-1)} \rangle}
{\vert \psi_{j_n}^{(n-1)} \vert}.
\label{eq:23}
\end{eqnarray}
\item The procedure finishes when the number of selected elements of the basis
set equals~$N_b$ given by Eq.~(\ref{eq:17}).
\end{itemize}

Finally, the corresponding Hamiltonian matrix is computed in the basis set of
scar functions, or alternatively in the equivalent basis set of auxiliary
functions, with the help of the wavelets, which provide an accuracy of at 
least 14 decimal places for the matrix elements. 
The diagonalization using standard routines~\cite{NR96} gives~$N_b$ 
eigenstates in the energy window defined in (\ref{eq:15}).
%
\subsection{\emph{Local} representation, scar intensities,
and participation ratio}
\label{sec:local_basis}

In order to analyze the performance of our method we will use in this work a
\emph{local} representation, $\vert \varphi_j^\loc\rangle$, defined as
%
\begin{equation}
\vert N \rangle=\sum_j^{N_b} C_{Nj} \vert \varphi_j^\loc \rangle,
\label{eq:25}
\end{equation}
where the coefficients $C_{Nj}=\langle \varphi_j^\loc \vert N \rangle$,
to reconstruct the eigenfunctions of the system in such a way that we can regain
the attractive initial intuitive interpretation.
Again, the procedure to compute the different $\vert \varphi_j^\loc\rangle$ functions
will be described here in an algorithmic way.
\begin{itemize}
\item \textbf{1.} The first element of the \emph{local} representation is taken as the
scar function, $\vert \psi_{j_1}^{(0)} \rangle$, that has the largest value of the scar
intensity, $x_1$, defined as
%
\begin{equation}
x_j^{(n)} = \vert \langle \psi_j^{(n)} \vert N \rangle \vert^2.
\label{eq:26}
\end{equation}
That is
%
\begin{equation}
\vert \varphi_1^\loc \rangle= \vert \psi_{j_1}^{(0)} \rangle, \;
\with \; x_1 \equiv x_{j_1}^{(0)}=\max\{x_j^{(0)}\},
\label{eq:27}
\end{equation}
\item \textbf{2.a} In order to identify the second largest scar intensity,
one has to calculate the orthogonal part of the remaining scar functions
$\vert \psi_{j}^{(0)} \rangle$ to $\vert \varphi_1^\loc \rangle$ by computing
%
\begin{equation}
\vert \psi_{j}^{(1)} \rangle=\vert \psi_{j}^{(0)} \rangle-\langle \varphi_1^\loc
\vert \psi_{j}^{(0)} \rangle \vert \varphi_1^\loc \rangle, \quad j \ne j_1.
\label{eq:28}
\end{equation}
\item \textbf{2.b} The second element of the \emph{local} representation is taken as
%
\begin{equation}
\vert \varphi_2^\loc \rangle= \frac{\vert \psi_{j_2}^{(1)} \rangle}
{\vert \psi_{j_2}^{(1)} \vert},
\label{eq:29}
\end{equation}
with $x_2\equiv x_{j_2}^{(1)}=\max\{x_j^{(1)}, j \ne j_1\}$.

The procedure is then continued, so that the $n$-th element of the \emph{local}
representation is computed in a similar way:
\item \textbf{n.a} The orthogonal part of the functions
$\vert \psi_{j}^{(n-2)} \rangle_{j \ne j_1, j_2,...,j_{n-1}}$ to
$\vert \varphi_{n-1}^\loc \rangle$
is calculated by computing
%
\begin{equation}
\vert \psi_{j}^{(n-1)} \rangle=\vert \psi_{j}^{(n-2)} \rangle-\langle \varphi_{n-1}^\loc
\vert \psi_{j}^{(n-2)} \rangle \vert \varphi_{n-1}^\loc \rangle ,
\label{eq:30}
\end{equation}
\item \textbf{n.b} and the $n$-th element of the \emph{local} representation is given by
%
\begin{equation}
\vert \varphi_n^\loc \rangle= \frac{\vert \psi_{j_n}^{(n-1)} \rangle}
{\vert \psi_{j_n}^{(n-1)} \vert},
\label{eq:31}
\end{equation}
with $x_n \equiv x_{j_n}^{(n-1)}=\max\{x_j^{(n-1)}, j \ne j_1,j_2,...,j_n\}$.
\end{itemize}

Let us remark here that the scar intensities,~$x_j$, provide information on the
localization properties of the eigenfunctions.
However, although the largest scar intensity,~$x_1$, provides a faithful
information on the localization on~$\vert \psi_{j_1}^{(0)} \rangle$,
the same is not true of~$\vert \psi_{j_n}^{(0)} \rangle$ since in their
construction, the contribution in the subspace spanned the functions defined
previously in the SGSM was subtracted.
Nevertheless, the sum~$x_1+x_2+...+x_n$ provides information about the projection of
$\vert N \rangle$ onto the subspace defined by
$\vert \psi_{j_1}^{(0)} \rangle, \vert \psi_{j_2}^{(0)} \rangle, \ldots,
\vert \psi_{j_n}^{(0)} \rangle$.

Let us finally present some useful results concerning the scar intensities
defined in Eq.~(\ref{eq:26}).
Assuming that the distribution of these magnitudes follows a Gaussian law
for chaotic eigenfunctions, a semiclassical approximation for the average
can be obtained, as discussed in Ref.~\onlinecite{ver07}.
Indeed, the averaged value of the $j$-th largest scar intensity is given by
%
\begin{equation}
\bar{x}_j \simeq \frac{1}{\bar{\sigma}_r} \left( \frac{2}{\pi} \right)^{1/2}
\left[\alpha_j-\ln\left(\alpha_j+\frac{9}{8}\right)+b_j+
\frac{b_j^2}{2}\right],
\label{eq:32}
\end{equation}
where
%
\begin{equation}
\bar{\sigma}_r=\rho \sigma_N, \;
\alpha_j \equiv \bar{z}_j+\ln\frac{\sqrt{2}\bar{\sigma}_r}{j}, \;
b_j \equiv \frac{\ln(\alpha_j+287/128)}{\alpha_j+17/8},
\label{eq:33}
\end{equation}
being $z$ a random variable with averages $\bar{z}_1 \simeq 0.577$
and $\bar{z}_2 \simeq 13/48$ for the two largest scar intensities.
Let us remark that $\bar{x}_j$ goes, respectively, to zero and infinity
for large and low values of the energy.
Also, the above defined semiclassical expression can be improved by including
a higher order correction term in~$\alpha_j$,
so that~$\alpha_j \equiv \bar{z}_j+\ln (\sqrt{2}\bar{\sigma}_r/j+c_j)$.
Adequate values for~$c_j$ are obtained by fitting to actual quantum
calculations of~$x_j$.
Notice that these corrections are of order~$\mathcal{O}(1/\bar{\sigma}_r)$,
while~$\alpha_j=\mathcal{O}(1/\ln \bar{\sigma}_r)$.

We conclude this subsection by considering the participation ratio, $\PR$,
that is defined, taking into account expression (\ref{eq:25}), as
\begin{equation}
\PR = \frac{\sum_{j=1}^{N_b} C_{Nj}^2}{\sum_{j=1}^{N_b} C_{Nj}^4}.
\label{eq:35}
\end{equation}
This magnitude gives an idea of the number of basis elements that are
approximately necessary to reconstruct the original eigenstate,
$\vert N \rangle$.
Accordingly, this is a parameter very relevant for our discussions,
since it can be used to compare the quality of two different basis sets.
Namely, the lower the value of the $\PR$, the better the basis. 
%
\section{Results}
\label{sec:results}
In this section we present and analyze our results on the use of scar
functions as basis sets for the calculation of eigenstates of
classically chaotic systems.
\subsection{Calculation of the eigenstates}
\label{subsec:calculation}

The lowest eigenvalues and eigenfunctions of the Hamiltonian operator
corresponding to the quartic oscillator defined in Eq.~(\ref{eq:1})
have been calculated using a basis set of scar functions constructed
using the SGSM defined above.

\begin{figure}
\centering
\includegraphics{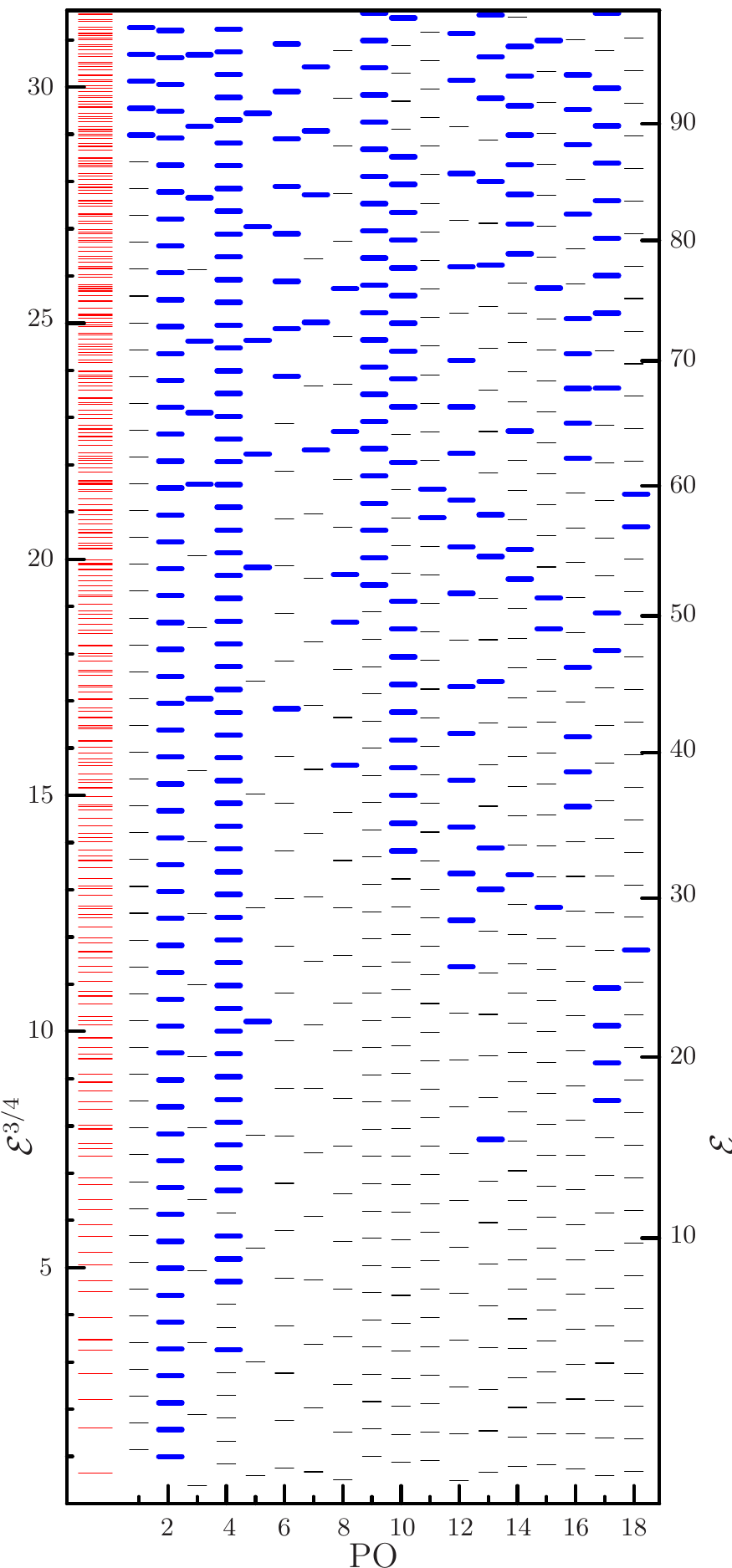}
\caption{(Color online) Scaled energies for the $A_1$ states of the
quartic oscillator (\ref{eq:1}):
(red leftmost tier) numerical eigenenergies,
(thin black lines) BS quantized energies,
(thick blue lines) energies of the scar functions selected by the SGSM.}
\label{fig:7}
\end{figure}

For this purpose, the $18$ POs presented in Fig.~\ref{fig:2},
chosen using the relevance parameter defined by Eq.~(\ref{eq:10}),
have been used.
From them, an initial basis set of scar functions was constructed,
defined by the reference energies
$\E^- = 0$ and $\E^+$=135, 150, 140, 140, and 82,
with $\bar{\sigma}$=0.5433, 0.5526, 0.5465, 0.5465, and 0.4622,
for the symmetry classes $A_1$, $A_2$, $B_1$, $B_2$, and $E$,
respectively.
This initial basis set consists of $\sim$900 scar functions for each
of the one dimensional symmetry classes: $A_1,A_2,B_1,B_2$,
and also for $E_1$ and $E_2$.
The associated energies were obtained from the BS quantization
rule (\ref{eq:9}) using the parameters given in
Tables~\ref{Table:I}-\ref{Table:III}.
The lowest values corresponding to the $A_1$ symmetry class,
rescaled with a power of $3/4$ so that they appear equally spaced,
are represented with thin black lines in Fig.~\ref{fig:7}.
%
The next step in our procedure is the reduction of this initial basis set.
This is done by applying the SGSM described in the Sect.~\ref{sec:GS},
using a value of $c_b=2$ in Eq.~(\ref{eq:17}).
With this criterion, $\sim$420 scar functions of each one dimensional
and $E_1$ and $E_2$ symmetry classes are automatically selected,
out of the whole set.
The energies corresponding to these selected $A_1$ scar functions
have been highlighted with thick blue lines in Fig.~\ref{fig:7}.

Using the resulting final basis set, we have computed the corresponding
Hamiltonian matrix and, by direct diagonalization, the eigenenergies and
eigenfunctions of the system.
The results for the 270 low-lying states of $A_1$ symmetry are shown
in the leftmost (red lines) tier of Fig.~\ref{fig:7}.
In Table~\ref{table:IIIbis} the 50 low-lying numerical values for all
symmetries are reported.
%
%
\begin{table}
\begin{tabular}{cccccc}
\hline
\hline
\multicolumn{1}{c}{$N$} & $A_1$ & $A_2$ & $B_1$ & $B_2$ & $E$ \\
\hline
\multicolumn{1}{c}{1} & 0.56323 & 4.1023 & 1.6175 & 2.5230 & 1.2241
\\
\multicolumn{1}{c}{2} & 1.8848 & 5.9503 & 2.7455 & 4.7256 & 2.2570
\\
\multicolumn{1}{c}{3} & 2.8638 & 7.5442 & 3.8675 & 6.0561 & 3.2537
\\
\multicolumn{1}{c}{4} & 3.8563 & 8.9638 & 5.0139 & 7.2567 & 3.6376
\\
\multicolumn{1}{c}{5} & 4.8286 & 9.7951 & 6.1773 & 8.2227 & 4.4506
\\
\multicolumn{1}{c}{6} & 5.2584 & 10.658 & 6.8364 & 9.1029 & 5.1299
\\
\multicolumn{1}{c}{7} & 6.2126 & 11.985 & 7.4816 & 10.468 & 5.5775
\\
\multicolumn{1}{c}{8} & 7.4115 & 12.861 & 8.7094 & 11.352 & 6.2642
\\
\multicolumn{1}{c}{9} & 7.9052 & 13.616 & 9.3168 & 11.914 & 6.8080
\\
\multicolumn{1}{c}{10} & 8.6947 & 14.793 & 10.080 & 12.708 & 6.9510\\
\multicolumn{1}{c}{11} & 9.3055 & 15.393 & 11.188 & 13.637 & 7.9451
\\
\multicolumn{1}{c}{12} & 10.087 & 16.069 & 11.534 & 14.276 & 8.1741
\\
\multicolumn{1}{c}{13} & 10.664 & 16.731 & 12.525 & 15.134 & 8.3392
\\
\multicolumn{1}{c}{14} & 11.452 & 17.702 & 12.908 & 15.928 & 9.3842
\\
\multicolumn{1}{c}{15} & 11.960 & 18.287 & 13.550 & 16.571 & 9.4158
\\
\multicolumn{1}{c}{16} & 12.790 & 19.115 & 14.309 & 17.215 & 9.8674
\\
\multicolumn{1}{c}{17} & 13.140 & 20.011 & 14.934 & 18.242 & 10.448
\\
\multicolumn{1}{c}{18} & 14.298 & 20.467 & 15.827 & 18.702 & 10.804
\\
\multicolumn{1}{c}{19} & 14.714 & 21.183 & 16.234 & 18.976 & 11.013
\\
\multicolumn{1}{c}{20} & 15.003 & 21.789 & 17.007 & 19.988 & 11.520
\\
\multicolumn{1}{c}{21} & 15.831 & 22.345 & 17.391 & 20.220 & 12.004
\\
\multicolumn{1}{c}{22} & 16.024 & 23.217 & 18.099 & 21.270 & 12.267
\\
\multicolumn{1}{c}{23} & 16.924 & 23.673 & 18.928 & 21.821 & 12.786
\\
\multicolumn{1}{c}{24} & 17.384 & 24.062 & 19.068 & 22.073 & 13.091
\\
\multicolumn{1}{c}{25} & 17.989 & 24.810 & 19.885 & 23.038 & 13.526
\\
\multicolumn{1}{c}{26} & 18.517 & 25.348 & 20.202 & 23.556 & 13.814
\\
\multicolumn{1}{c}{27} & 18.972 & 26.495 & 20.604 & 24.428 & 14.109
\\
\multicolumn{1}{c}{28} & 19.905 & 26.629 & 21.668 & 24.591 & 14.541
\\
\multicolumn{1}{c}{29} & 20.184 & 27.083 & 22.118 & 25.140 & 14.729
\\
\multicolumn{1}{c}{30} & 20.592 & 27.860 & 22.358 & 25.967 & 15.069
\\
\multicolumn{1}{c}{31} & 21.163 & 28.462 & 23.087 & 26.104 & 15.646
\\
\multicolumn{1}{c}{32} & 21.931 & 28.946 & 23.754 & 26.682 & 15.893
\\
\multicolumn{1}{c}{33} & 22.228 & 29.614 & 24.066 & 27.357 & 16.209
\\
\multicolumn{1}{c}{34} & 22.458 & 29.711 & 24.939 & 27.950 & 16.600
\\
\multicolumn{1}{c}{35} & 23.247 & 30.334 & 25.079 & 28.478 & 16.801
\\
\multicolumn{1}{c}{36} & 23.727 & 31.099 & 25.513 & 29.093 & 17.373
\\
\multicolumn{1}{c}{37} & 24.010 & 31.516 & 25.916 & 29.241 & 17.407
\\
\multicolumn{1}{c}{38} & 24.653 & 32.199 & 26.817 & 30.233 & 17.768
\\
\multicolumn{1}{c}{39} & 25.231 & 32.626 & 27.121 & 30.333 & 18.152
\\
\multicolumn{1}{c}{40} & 25.576 & 32.831 & 27.443 & 31.015 & 18.339
\\
\multicolumn{1}{c}{41} & 26.121 & 33.681 & 28.180 & 31.511 & 18.764
\\
\multicolumn{1}{c}{42} & 26.530 & 34.089 & 28.491 & 32.184 & 19.147
\\
\multicolumn{1}{c}{43} & 27.098 & 34.516 & 28.932 & 32.312 & 19.174
\\
\multicolumn{1}{c}{44} & 27.403 & 35.335 & 29.345 & 33.214 & 19.765
\\
\multicolumn{1}{c}{45} & 28.097 & 35.928 & 29.980 & 33.441 & 20.005
\\
\multicolumn{1}{c}{46} & 28.727 & 36.002 & 30.711 & 34.020 & 20.136
\\
\multicolumn{1}{c}{47} & 28.946 & 36.551 & 31.076 & 34.323 & 20.642
\\
\multicolumn{1}{c}{48} & 29.342 & 36.933 & 31.467 & 34.651 & 20.898
\\
\multicolumn{1}{c}{49} & 29.475 & 37.373 & 31.568 & 34.953 & 21.120
\\
\multicolumn{1}{c}{50} & 30.225 & 37.855 & 32.483 & 35.880 & 21.440
\\
\hline
\
\end{tabular}
\caption{Eigenenergies, $\E$, for the eigenstates
of the quartic oscillator (\ref{eq:1}) obtained with our basis set
of scar functions.}
\label{table:IIIbis}
\end{table}
Several comments are in order.
First, a careful comparison reveals that the first $\sim$400
of each symmetry have the same accuracy 
(not only the eigenenergies
but also the eigenfunctions) that is obtained with other standard 
methods~\cite{pullenedmonds81_carne84_EHP89},
and then the number of well converged states in our calculation  
is of the same order as the number of elements in the basis set.
Second, some first qualitative conclusions regarding the excellent performance
of our basis set can be obtained from a careful consideration of the results
presented in Fig.~\ref{fig:7}.
To this end, recall again the extremely low dispersion of our scar functions,
which means that they minimally spread among (or contribute to) states
in the eigenenergy spectrum far from their BS quantized energies.
For example, for the most excited states only~$\sim$6 scar basis
functions are needed for a satisfactory description.
This number dramatically decreases for smaller excitations,
getting as low as~$\sim$1 for states in the interval
$\Delta \E \sim 0-5$,~$\sim$2 for states in the interval~$\Delta \E \sim 5-10$,
and so on as the scar states get ``bright'', being highlighted in blue in our plot.
This argument will be made more quantitative in the rest of the discussions
presented below in this section, and particularly when the eigenstate 
participation ratios are considered.
Finally, it is worth emphasizing that besides the type of calculation
presented here, namely, the computation of the low-lying eigenstates,
the SGSM introduced in this work can be advantageously used in the
computation of eigenstates in an energy window, i.e.~$\E^- \neq 0$,
something which is specially useful to compute only highly excited
eigenfunctions with small basis sets.
This is something that we have tested for different energy ranges.
In the next subsection we present the results associated to the 
eigenfunctions $\vert N \rangle_{A_1}$ with $N=271-282$, 
calculated restricting the calculation to the energy window 
$100 < \E < 103$ with a basis set formed by only 25 scar functions. 
It is quite impressive that using such a small basis set the error in the 
energy is smaller than $0.12$ in units of the mean level spacing 
(actually, it is even smaller than $0.06$ for all the computed 
eigenvalues except for $\vert 279 \rangle_{A_1}$), whereas 
the overlap of all the computed eigenfunctions with the 
corresponding exact ones exceeds $99.8\%$. 
Further details on these results will be presented in the 
following section.

\subsection{Reconstruction of the eigenfunctions}
\label{subsec.spectrum_scar}

Let us now analyze the results obtained in the previous subsection.
In the first place, we discuss the structure of some representative examples
of the eigenfunctions obtained for the quartic oscillator (\ref{eq:1})
with our basis set of scar functions, by examining their reconstruction
using the \emph{local} representation described in subsection
\ref{sec:local_basis}.

%
\begin{figure}
\includegraphics{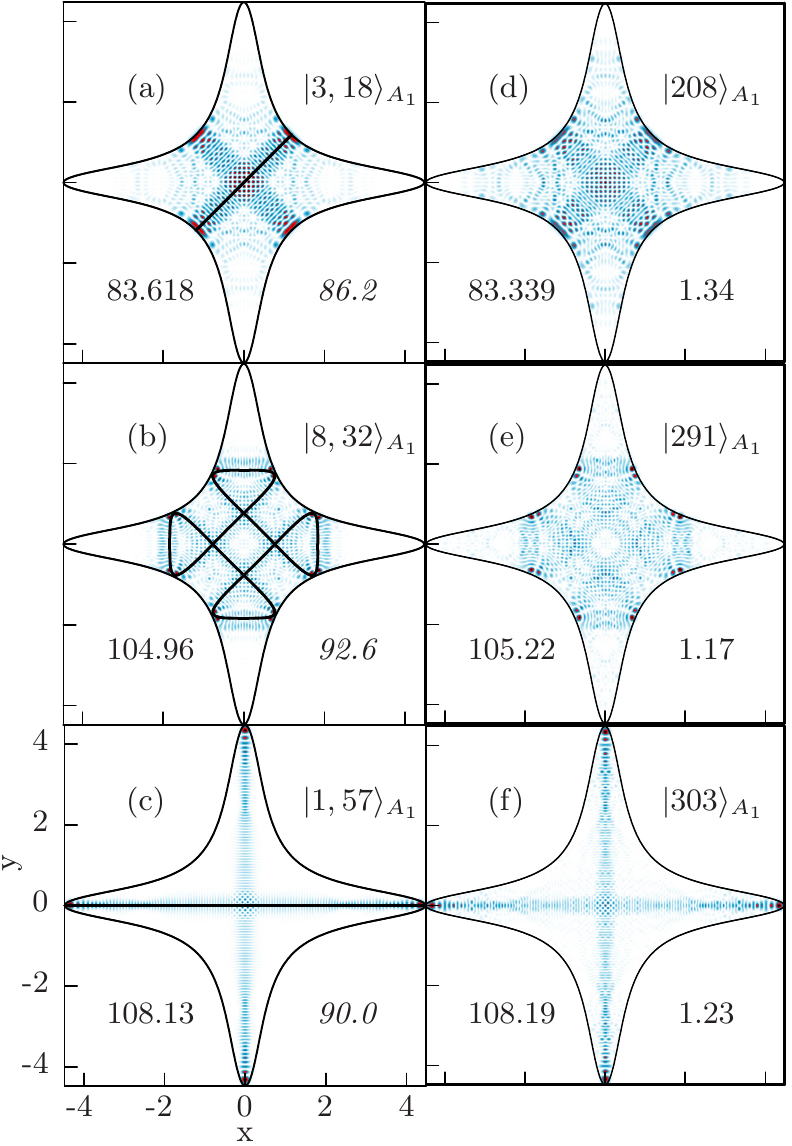}
\caption{(Color online)
(Right tier) Probability density for the eigenfunctions $\vert 208 \rangle_{A_1}$,
$\vert 291 \rangle_{A_1}$, and $\vert 303 \rangle_{A_1}$ obtained from our
variational calculation using a basis set of scar functions,
and (left tier) the same functions reconstructed using only one scar function
$\vert{\rm PO}, n \rangle_{A_1}$ of the basis set.
The scarring PO has been plotted superimposed in panels (a)-(c).
The energy is shown on the left corner of each panel, and the overlaps
of the eigenfunctions with the scar functions (left tier) and the participation
ratios (right tier) are given on the right corner of the panels.}
\label{fig:8}
\end{figure}
We start by the simplest case, which corresponds to states with
eigenfunctions that appear strongly scarred in the sense discussed
by Heller in Ref.~\onlinecite{heller84}. This happens, for
example, for states $\vert 208 \rangle_{A_1}$, $\vert 291
\rangle_{A_1}$, and $\vert 303 \rangle_{A_1}$, which are highly
localized over POs numbers 3, 8, and 1 of Fig.~\ref{fig:2},
respectively. The probability densities are shown in the right
tier of Fig.~\ref{fig:8}. In all cases, the first, and almost
exclusively contributing, element of the reconstruction is a scar
function corresponding to the PO scarring the eigenfunction, as
could be expected \emph{a priori}. These scar functions, which
will be labelled $\vert {\rm PO}, n \rangle_\chi$ with PO
indicating the number of the orbit in Fig.~\ref{fig:2}, $n$ the
number of excitations along it corresponding to the BS
quantization condition (\ref{eq:9}), and $\chi$ the IR, are
presented in the left tier of the figure. The PO has been plotted
superimposed to the corresponding probability density. The
associated energies are given in the lower left corner of each
panel. In the lower right corner of the panels (a)-(c), we have
indicated the value of the overlap between the eigenstate and the
basis set scar function. As can be seen, this overlap is always
larger than 86.0\% in all cases considered here. Furthermore, the
fact that these states are well represented by only one state of
the scar basis set makes the values of the corresponding participation 
ratio very small, as discussed in the subsection \ref{subsec:pr}. 
The values of these participation ratios are given in the right 
corners of panels (d)-(f) in Fig.~\ref{fig:8}.

%
\begin{figure}
\includegraphics{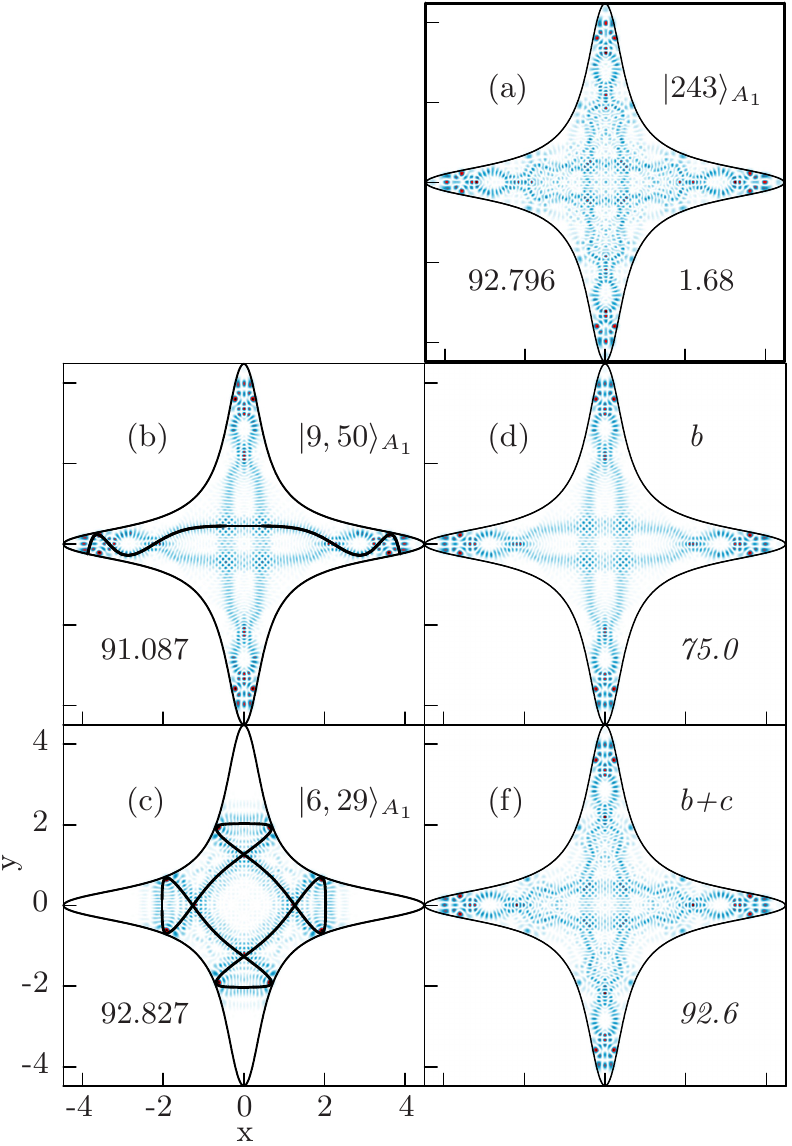}
\caption{(Color online)
Same as in Fig.~\ref{fig:8} for the eigenfunction $\vert 243 \rangle_{A_1}$ (a),
which requires two scar functions, $\vert 9,50 \rangle_{A_1}$ (b)
and $\vert 6,29 \rangle_{A_1}$ (c) for its reconstruction,
which is shown in (d) and (e).
The information included in the different panels is the same as in
Fig.~\ref{fig:8}.}
\label{fig:9}
\end{figure}
Let us consider next eigenstates with more complex structure.
This is the case, for example, of $\vert 243 \rangle_{A_1}$
shown at the top of Fig.~\ref{fig:9}.
As can be seen in panel~(a), the eigenfunction for this state is
concentrated on a single PO, i.e.~PO number~9, but the localization
is not as strong as in the previous examples.
Actually, the reconstruction of this eigenstate requires the combination
of at least two scar functions, as the results presented in the other panels
indicate.
Indeed, the scar basis function~$\vert 9,50 \rangle_{A_1}$, shown in panel~(b),
only accounts for~$75.0\%$ of the eigenstate, while when function
$\vert 6,29 \rangle_{A_1}$ [cf. panel (c) of Fig.~\ref{fig:9}] is included,
this figure raises to an acceptable~$92.6\%$.
As discussed before, the corresponding value of the participation ratio is
expected to be larger, $\PR=1.68$, in this case.

%
\begin{figure}
\includegraphics{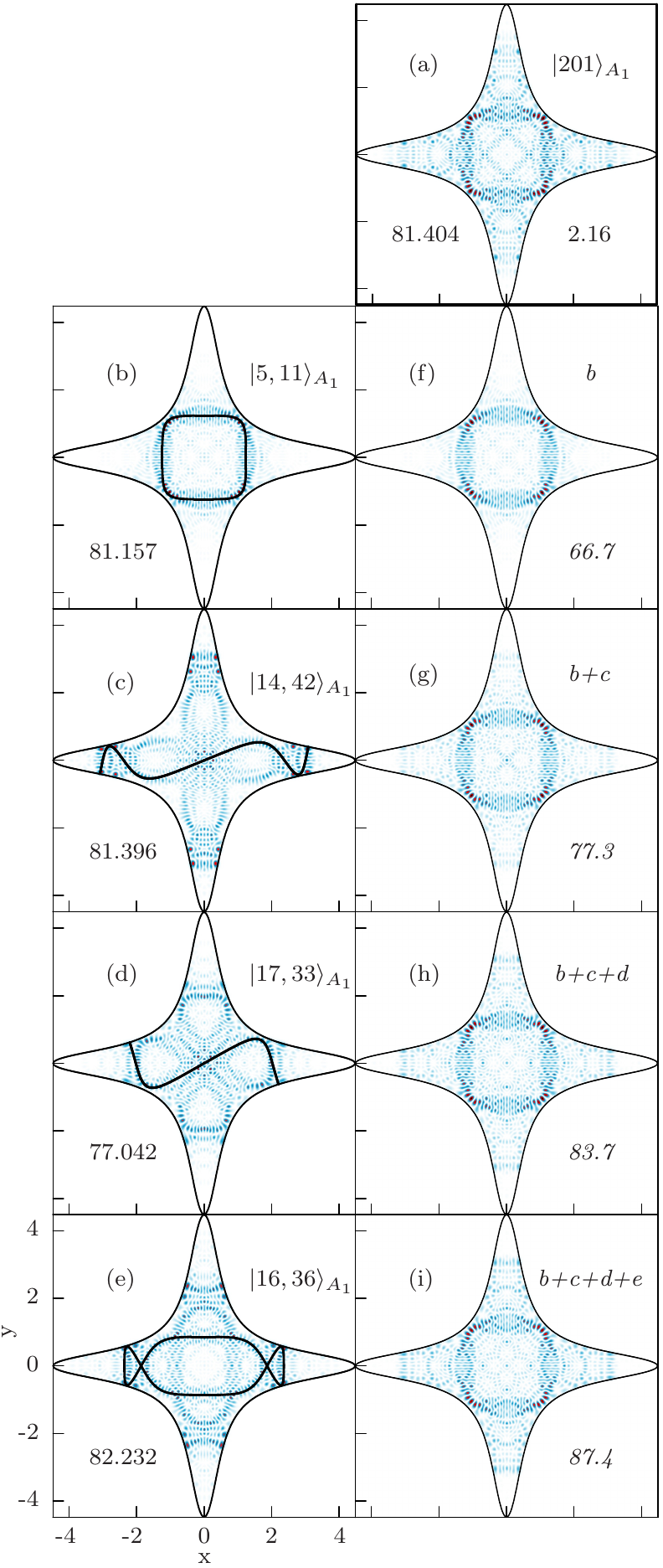}
\caption{Same as Fig.~\ref{fig:9} for $\vert 201 \rangle_{A_1}$
which requires four scar functions for its reconstruction.}
\label{fig:10}
\end{figure}
A similar example, but with an eigenfunction exhibiting an even
more complicated structure is shown in Fig.~\ref{fig:10}.
In this case, the eigenfunction $\vert 201 \rangle_{A_1}$,
which is localized over the ``box'' PO number 5 [see panel (a)],
requires the combination of at least four scar functions localized
over different POs [see panels (b)-(e)] for an adequate reconstruction.
The relevant figures for this reconstruction, i.e.~overlaps and participation ratios,
are indicated in the figure, and the same comments made before
apply to this case.

%
\begin{figure}
\includegraphics{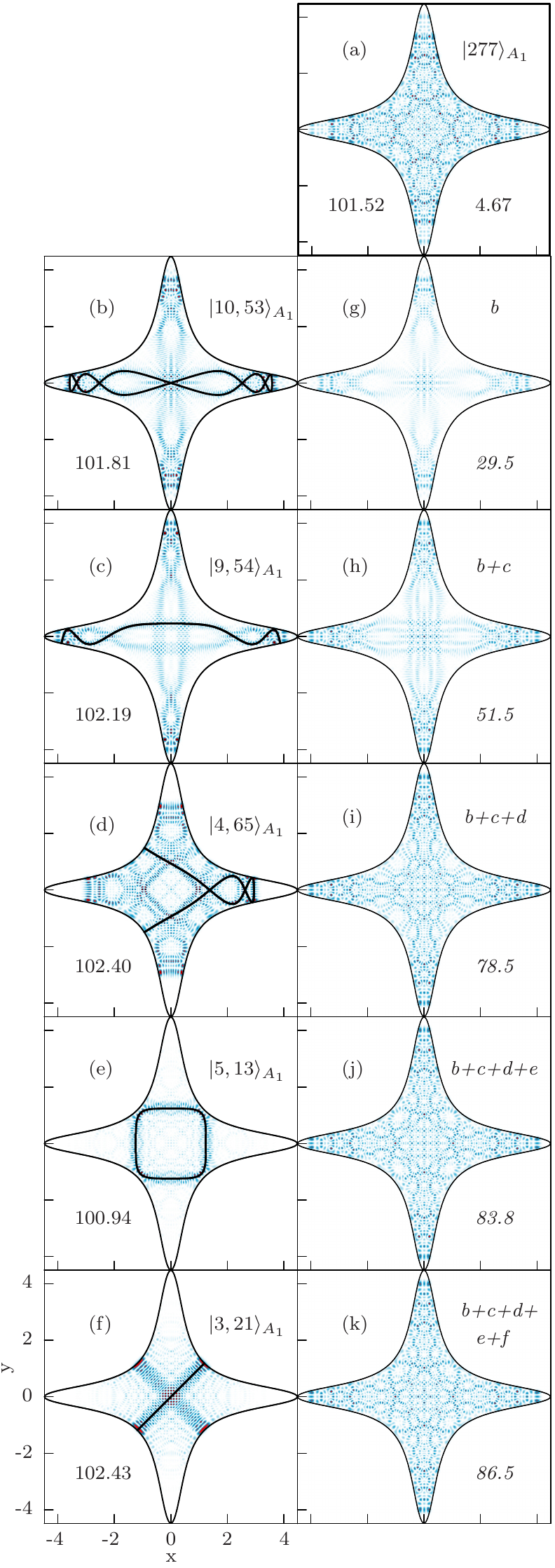}
\caption{Same as Fig.~\ref{fig:9} for $\vert 277 \rangle_{A_1}$.}
\label{fig:11}
\end{figure}
%
\begin{figure}
\includegraphics{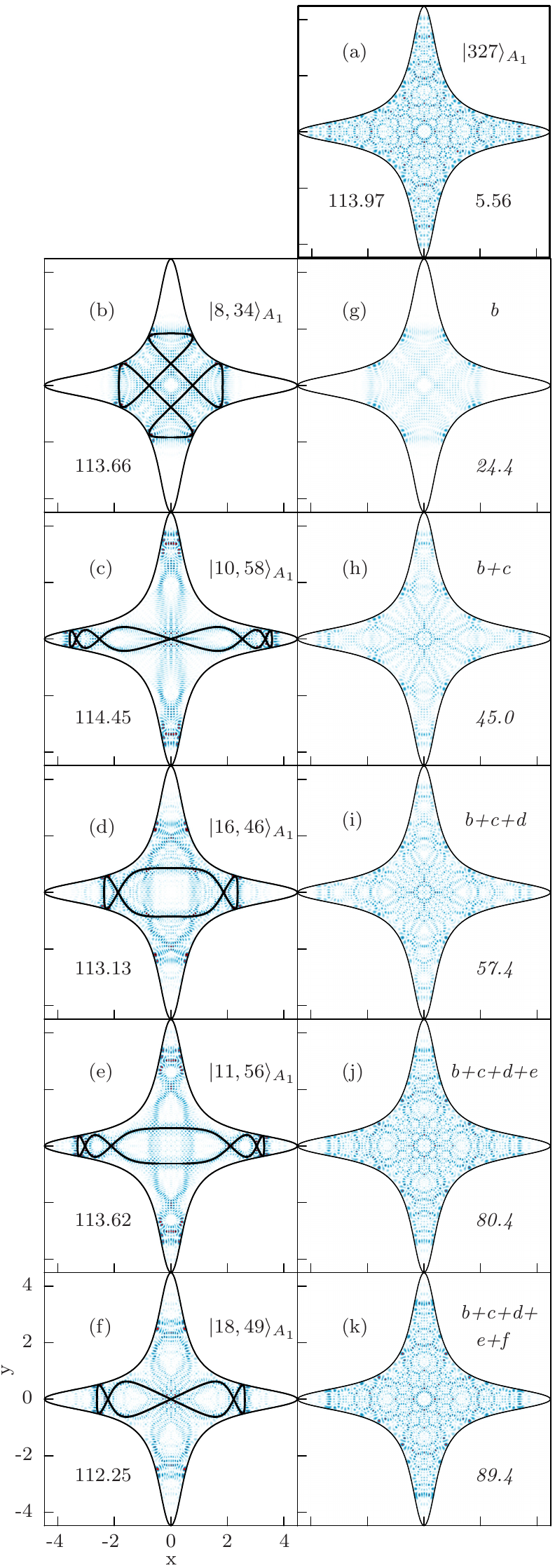}
\caption{Same as Fig.~\ref{fig:9} for $\vert 327 \rangle_{A_1}$.}
\label{fig:12}
\end{figure}
To conclude this series of examples, let us consider finally the most
general case consisting of (non-scarred) eigenstates showing the very
irregular patterns characteristic of chaotic states~\cite{berry77}.
For this purpose, we present in Figs.~\ref{fig:11} (a) and \ref{fig:12} (a)
the probability density for states $\vert 277 \rangle_{A_1}$
and $\vert 327 \rangle_{A_1}$, respectively.
The complex nodal structure inherent to these wavefunctions can be unfolded,
however, by our SGSM which reveals the importance of each PO,
thus proving a dynamically oriented analysis of them.
As can be seen, the two cases considered here are essentially reconstructed
by combining only five scar functions, which are also shown in order of
importance in the two figures [panels (b)-(f)].

From the data contained in the figures, notice how in all the examples 
presented here the scar functions giving the largest contributions to the 
reconstruction of a given eigenstate are those whose BS quantized energies 
are closer that of the corresponding eigenenergy.

\begin{figure}
\includegraphics{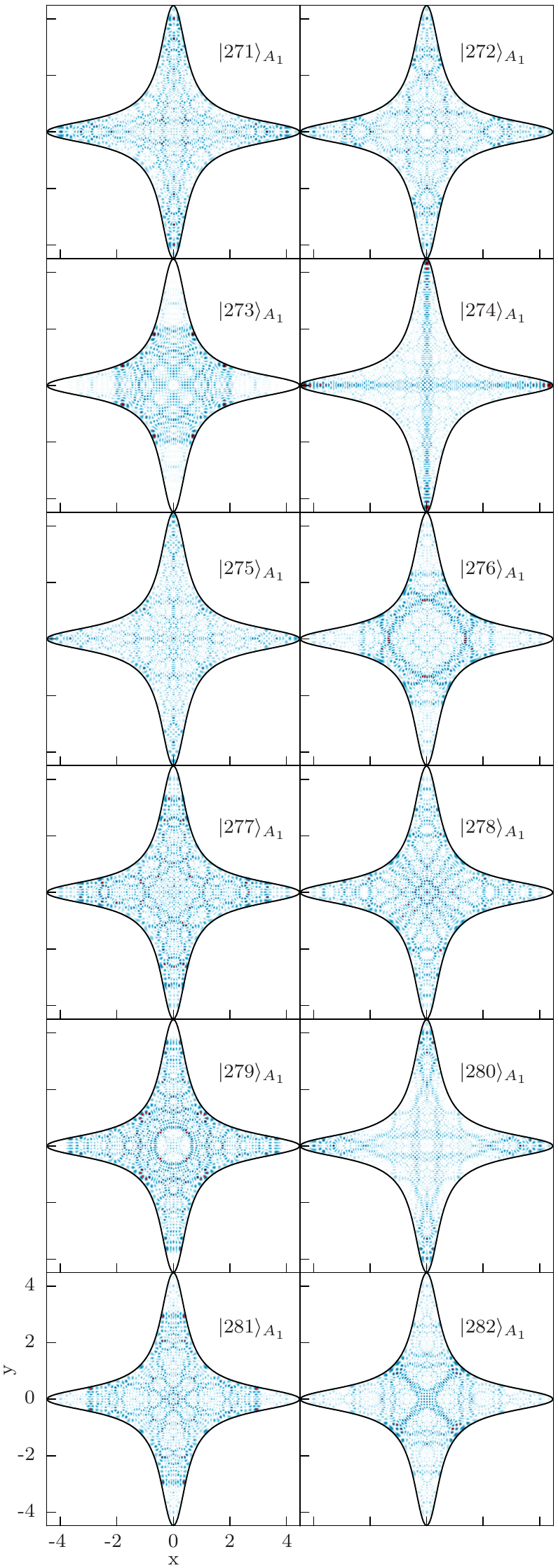}
\caption{(Color online) Quartic oscillator eigenfunctions $\vert N \rangle_{A_1}$
corresponding to the reconstruction data of Table~\ref{Table:window}.
Same coordinates and scaling as in Fig.~\ref{fig:5}.}
\label{fig:window}
\end{figure}

We close this subsection by presenting in 
Table~\ref{Table:window} the structure of the eigenfunctions 
calculated in the energy window $100<\E<103$. 
These eigenfunctions are shown in Fig.~\ref{fig:window} .
As in the previous examples, most of the probability density is 
reconstructed by combining few basis elements, 
thus demonstrating the efficiency of 
our method for the calculation of excited states.

Further results on the structure of the eigenfunctions of the quartic oscillator in
our scar function basis set are systematically presented in the supplemental material~\cite{supp}.
%
\begin{table*}
\caption{Reconstruction of the eigenstates $\vert N \rangle_{A_1}$ 
calculated in the energy window $100<\E<103$ with the 25 scar functions 
$\vert {\rm PO}, n \rangle_{A_1}$.
The participation ratio, $\PR$, defined in Eq.~(\ref{eq:35}) and the 
accumulated reconstruction overlap, $\Sigma\%$, are also given.}
\begin{tabular}{rrrrrrrrrrrr}
\hline\hline
$N$ & $\PR$ & PO,$n$,$\Sigma\%$ & & & & & & & & & \\
\hline
271 & 4.81 & 9,53,40.5 & 9,53,40.5 & 4,64,56.8 & 8,31,62.6 & 18,45,69.5 & 14,49,74.0 & 15,47,77.9 & 5,13,81.4 & 10,52,84.5 & 17,39,87.4 \\
272 & 3.53 & 15,47,48.4 & 13,35,62.2 & 9,53,76.7 & 10,52,82.4 & 4,64,86.5 &  &  &  &  &  \\
273 & 2.09 & 8,31,67.6 & 6,31,76.6 & 5,13,87.1 &  &  &  &  &  &  &  \\ 
274 & 2.00 & 1,54,69.7 & 5,13,78.2 & 8,31,84.0 & 18,45,88.1 &  &  &  &  &  &  \\ 
275 & 5.24 & 2,54,31.0 & 5,13,55.9 & 8,31,69.0 & 13,35,76.0 & 4,64,84.9 & 17,39,87.5 &  &  &  &  \\ 
276 & 3.36 & 6,31,28.7 & 8,31,72.6 & 10,53,85.5 &  &  &  &  &  &  &  \\ 
277 & 4.68 & 10,53,29.5 & 9,54,51.5 & 4,65,78.5 & 5,13,83.8 & 3,21,86.5 &  &  &  &  &  \\
278 & 5.27 & 10,53,30.8 & 3,21,53.3 & 4,65,71.4 & 17,40,78.2 & 18,46,82.0 & 6,31,84.6 & 8,31,86.3 &  &  &  \\ 
279 & 7.85 & 9,54,21.8 & 4,65,34.2 & 15,48,40.1 & 3,21,45.5 & 13,36,52.5 & 5,13,56.1 & 2,55,58.5 & 1,55,79.4 & 1,54,81.1 & 17,40,81.9  \\
       &         & 18,46,88.1 &  &  &  &  &  &  &  &  \\
280 & 2.08 & 9,54,67.7 & 15,48,71.3 & 16,43,73.9 & 10,53,76.5 & 5,13,77.5 & 18,46,78.3 & 13,36,79.2 & 2,55,79.9 & 1,55,92.6 &  \\
281 & 3.30 & 17,40,49.3 & 15,48,69.7 & 9,54,77.9 & 4,65,85.4 &  &  &  &  &  &  \\ 
282 & 4.20 & 3,21,44.3 & 13,36,52.7 & 15,48,65.7 & 11,52,73.7 & 4,65,82.0 & 10,53,84.1 & 9,54,86.1 &  &  &  \\
\hline
\end{tabular}
\label{Table:window}
\end{table*}
%
\subsection{Scar intensities}
\label{subsec:intens}
%
\begin{figure}
\includegraphics[width=\columnwidth,angle=0]{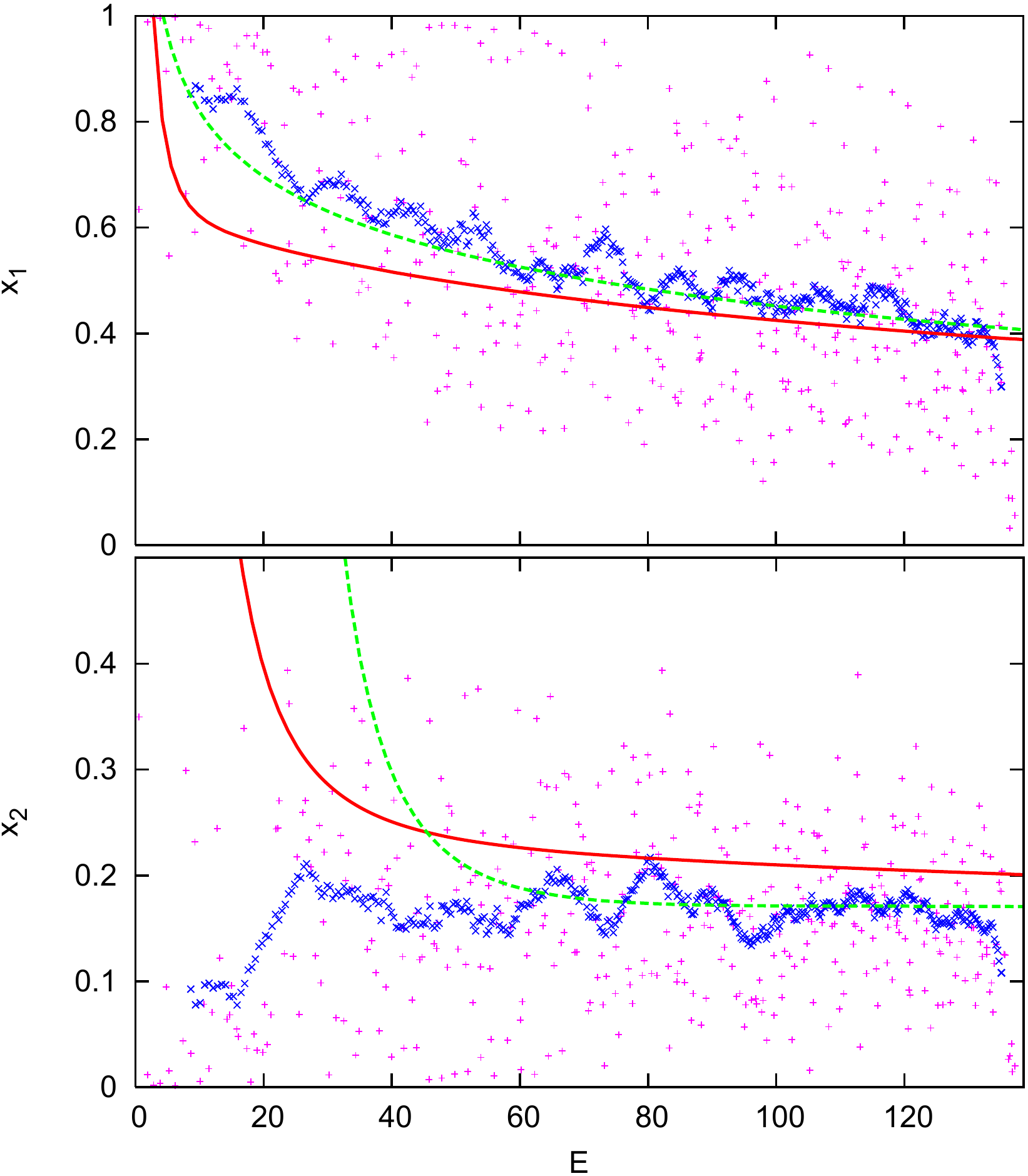}
\caption{(Color online)
Largest scar intensities $x_1$ (top) and $x_2$ (bottom) (pink plus signs)
for the eigenfunction with $A_1$ symmetry obtained in our calculation
with a basis set of scar functions.
The average value, computed as a mobile mean, is plotted with the superimposed
with blue crosses.
The semiclassical estimates obtained with Eq.~(\ref{eq:32}) 
for $c_1 = c_2 =0$ and $c_1 \simeq -c_2 \simeq 0.30$ are also plotted with 
red solid and green dashed lines, respectively.}
\label{fig:16}
\end{figure}
A more global idea of how the eigenfunctions are reconstructed by the scar
functions can be obtained by considering the corresponding scar intensities
defined in Eq.~(\ref{eq:26}).
In the two panels of Fig.~\ref{fig:16} we show (pink plus signs)
the variation with the energy of the two largest scar intensities,~$x_1$ and~$x_2$,
for the computed~$A_1$ eigenfunctions.
Recall that these quantities give the (squared) contribution of the two most
important scar functions in the \emph{local} representation of each eigenfunction.
As can be seen, both~$x_1$ and~$x_2$ fluctuate violently,
thus appearing rather dispersed in the figure.
To clearly observe the tendency in the data we have also plotted the
corresponding mobile mean (blue crosses), computed using~20 points.
The results indicate that the largest intensity,~$x_1$, starts from
a very high value, and then monotonically decreases,
first very rapidly up to~$\E \sim 22$, and then much slower,
never getting in mean below~$\sim$0.4
in the energy range that we are considering.
Notice that the points where~$x_1$ is much larger that its average value
correspond to scarred states in the sense of Heller~\cite{heller84}.
Simultaneously, the second largest contribution,~$x_2$,
first increases up to~$\sim$0.2 at~$\E \sim 22$, and then decreases
extremely slowly (notice that although this behavior in not noticeable
in the plot this is true since~$x_1>x_2$).
Moreover, the values given by the semiclassical approximations~(\ref{eq:32}),
without and with the higher order correction terms with~$c_1 \simeq -c_2 \simeq 0.30$,
have also been plotted superimposed in red solid and green dashed lines,
respectively.
The agreement between the quantum and semiclassical calculations,
particularly in the second case (green dashed line) is rather good,
except for low values of the energy due to the (unrealistic)
singularity inherent to Eq.~(\ref{eq:32}).

\begin{figure}
\includegraphics[width=0.6\columnwidth,angle=270]{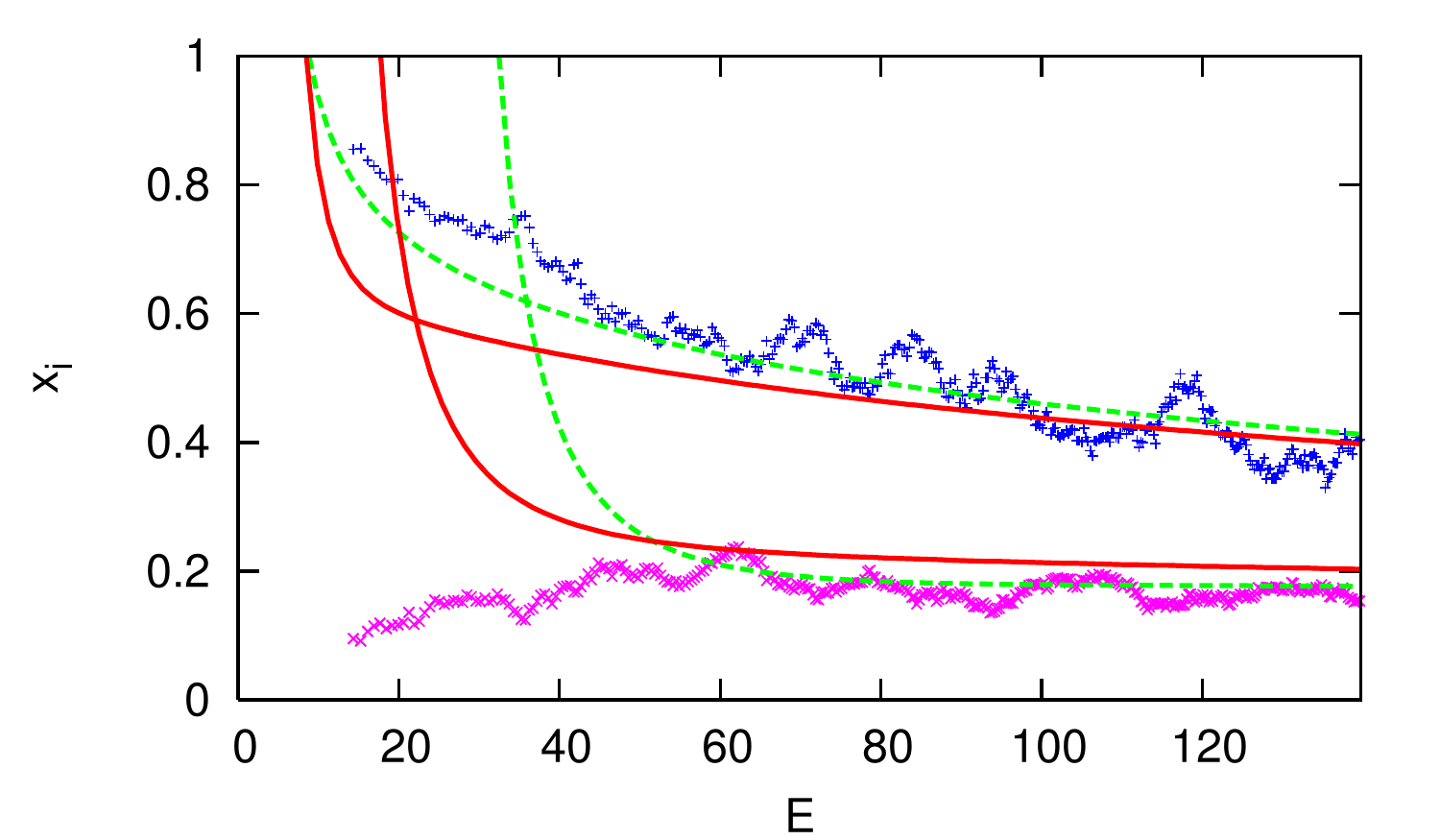}
\caption{(Color online)
Average value of the largest scar intensities $x_1$ (blue upper plus signs)
and $x_2$ (pink lower crosses) of the eigenfunctions of $A_2$ symmetry.
The semiclassical estimates obtained with Eq.~(\ref{eq:32})
are also plotted with red solid and green dashed lines, respectively.}
\label{fig:17}
\end{figure}

Similar results for the $A_2$ eigenstates are shown in Fig.~\ref{fig:17}.
Here, only the values of the averages and the semiclassical estimates
are shown for simplicity.
The same comments made before for the $A_1$ symmetry apply.
%

\subsection{Participation ratios}
\label{subsec:pr}

Also interesting for the analysis of the eigenstate structure is the
consideration of the participation ratios discussed in subsection \ref{sec:local_basis}.
To give an idea of the expected bounds for these variables
let us consider two extreme cases.
First, the minimum value of the participation ratio is equal to $\PR_{\rm min}=1$,
which is obviously obtained when the system eigenfunctions are used.
On the other extreme, a large value of the participation ratio corresponds to a situation
when all basis functions significantly contribute to the description
of each eigenstate.
One such case, that retains however the character and then the efficiency
of a semiclassical description, is that of Ref.~\onlinecite{bogo92}.
There, all calculations are done in a characteristic Poincar\'e SOS
of an ergodic system.
Accordingly: 1) the size of the basis set, $N_b$, can be estimated
by the Weyl expression $N_b=A_{tr}/(2\pi\hbar)$, being $A_{tr}$
the area in the SOS, and 2) all the coefficients
in a normalized expansion are expected to behave as random complex
independent variables following a Gaussian distribution of zero mean
and dispersion $1/N_{b}^{1/2}$.
Then, it is straightforward to show that
%
\begin{equation}
\PR_{\scc}=\frac{A_{tr}}{4\pi\hbar}.
\label{eq:43}
\end{equation}
Let us remark that $\PR_{\scc}$ is still a reasonably small value,
since it is associated to an optimized, in the semiclassical sense, basis set,
thus giving participation ratio values much smaller than, for example,
for other frequently used basis sets,
such as harmonic oscillator products~\cite{pullenedmonds81_carne84_EHP89},
or a discrete variable representation~\cite{dvr}.

%
\begin{figure}
\includegraphics[width=\columnwidth,angle=0]{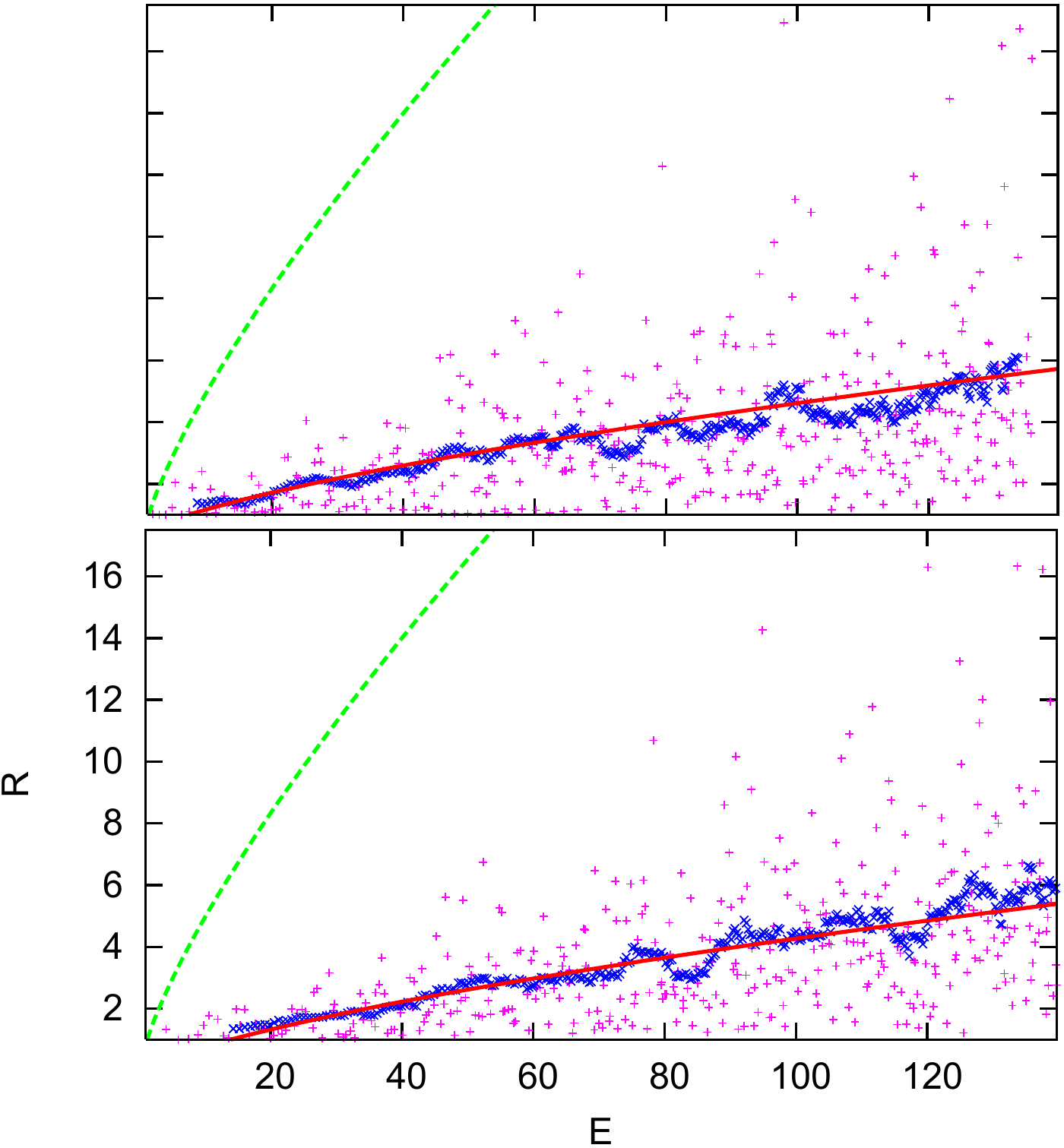}
\caption{(Color online)
Participation ratio in the basis set of scar functions (pink crosses)
for the eigenfunctions of $A_1$ (top) and $A_2$ symmetry,
as a function of the energy.
The average, computed as a mobile mean, is shown with blue crosses,
and it is well approximated with the continuous red curve given by the
simple expression (\ref{eq:41}).
The dashed green line represents semiclassical estimate given by
Eq.~(\ref{eq:43}).
}
\label{fig:18}
\end{figure}
In Fig.~\ref{fig:18} we plot with pink plus signs these participation ratios in the
\emph{local} representation as a function of the energy for the
$A_1$ (top) and the $A_2$ (bottom) symmetry eigenfunctions, respectively.
Similarly to what happens with the scar intensities we find a
wildly oscillatory behavior.
Accordingly, we consider the averaged value,
computed again as a mobile mean, which is also shown in the graph
using blue crosses.
The results obtained with the semiclassical expression (\ref{eq:43})
have also been plotted for comparison with dashed green line.
As can be seen, for energies up to $\E \sim 22$ the values of the
participation ratios are very small, and always lower than 3.
As energy increases the mean participation ratio also increases, but very moderately,
and it can be accurately fitted with the simple expression
(also shown in the figure with red continuous line)
%
\begin{equation}
\bar{\PR} = \zeta \bar{\sigma}_r,
\label{eq:41}
\end{equation}
in terms of the variable $\bar{\sigma}_r$ defined in Eq.~(\ref{eq:33}).
Here $\zeta=8/3$ for the one dimensional symmetry representations,
$A$ and $B$, and $\zeta=2$ for the two dimensional one, $E$.
The moderate increment behavior found is an indication of the quality
and good performance of the scar basis set we have used in our calculation,
and shows how a given eigenstate can be reconstructed with just a
few number of basis functions.
Actually, it has been numerically checked that the cases with
the highest values of the participation ratio, this number can be substantially reduced
by increasing the size of the basis by including more POs.

\begin{figure}
\includegraphics[width=5cm,angle=270]{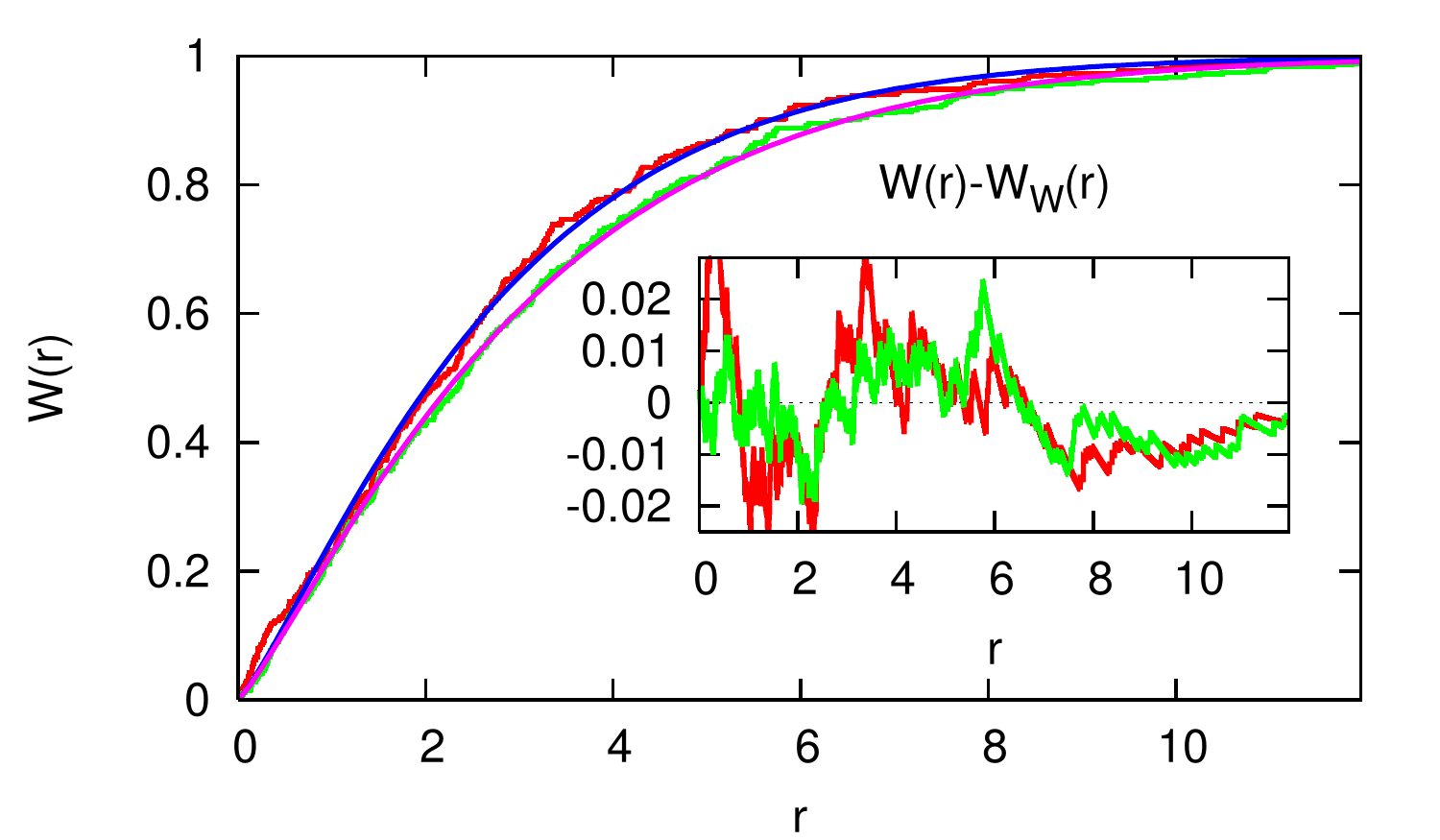}
\caption{(Color online)
Accumulated probability distribution, $W(\pr)$, of the participation ratio
adimensional coefficient, $\pr$, for the eigenfunctions of $A_1$ (red upper points)
and $A_2$ (green lower points) symmetry obtained in our calculation with a basis set
of scar functions.
The corresponding values of the Weibull distribution, $W_{\W}(\pr)$, for
$l_{A_1}=2.8323\pm 0.0003, k_{A_1}=1.2030 \pm 0.0002,
l_{A_2}=3.1867 \pm 0.0001, k_{A_2}=1.1753 \pm 0.0001$ are shown
in continuous blue and dashed pink lines, respectively.
The difference $W(\pr)-W_\W(\pr)$ is shown in the inset.}
\label{fig:19}
\end{figure}

The participation ratio results can be further analyzed by considering the associated
statistics.
For this purpose we define the adimensional coefficient
%
\begin{equation}
\pr=\frac{\PR-1}{\bar{\PR}},
\label{eq:44}
\end{equation}
which is a positive random variable that is found to follow
the Weibull distribution~\cite{weib51},
%
\begin{equation}
P_{\W}(x)=\frac{k}{l} \left( \frac{x}{l} \right)^{k-1} e^{-(x/l)^k},
\label{eq:45}
\end{equation}
$k$ and $l$ being fitting parameters.
The associated accumulated distribution can be easily calculated as
%
\begin{equation}
W_\W(\pr)=\int_0^{\pr} P_\W(x) \; dx = 1-e^{-(\pr/l)^k}.
\label{eq:46}
\end{equation}
The corresponding results for both the computed and fitted accumulated
distributions $W(\pr)$ and $W_\W(\pr)$ for the eigenfunctions of
$A_1$ and $A_2$ symmetry are shown in Fig.~\ref{fig:19}.
The difference between $W(\pr)$ and $W_{\W}(\pr)$, or error,
is also shown in the inset.
As can be seen, this difference is very small for both symmetries,
as it is also the case for the rest of the symmetry classes.
%
%
\subsection{Upper bounds to errors in eigenenergies and eigenfunctions}

\label{subsec:error}

In this subsection we obtain expressions for the upper bound to
the error in the eigenenergy and eigenfunctions obtained in our
calculation \emph{versus} their dispersion~\cite{ver08}. These
upper bounds are specially useful in the calculation of highly
excited states, when large basis sets are required.

Usually, the convergence of approximate eigenstates, both in energy and
wave functions, is assessed by comparing the results obtained using two
basis sets of different size.
In our case, we will take as the 'exact' results,
denoted by $\E'_N, \vert N' \rangle$,
those obtained variationally by diagonalization in a very large basis
set ($\sim$5000 elements) of harmonic oscillator eigenfunctions 
\cite{pullenedmonds81_carne84_EHP89,Waterland}.

\begin{figure}
\includegraphics[width=7cm,angle=270]{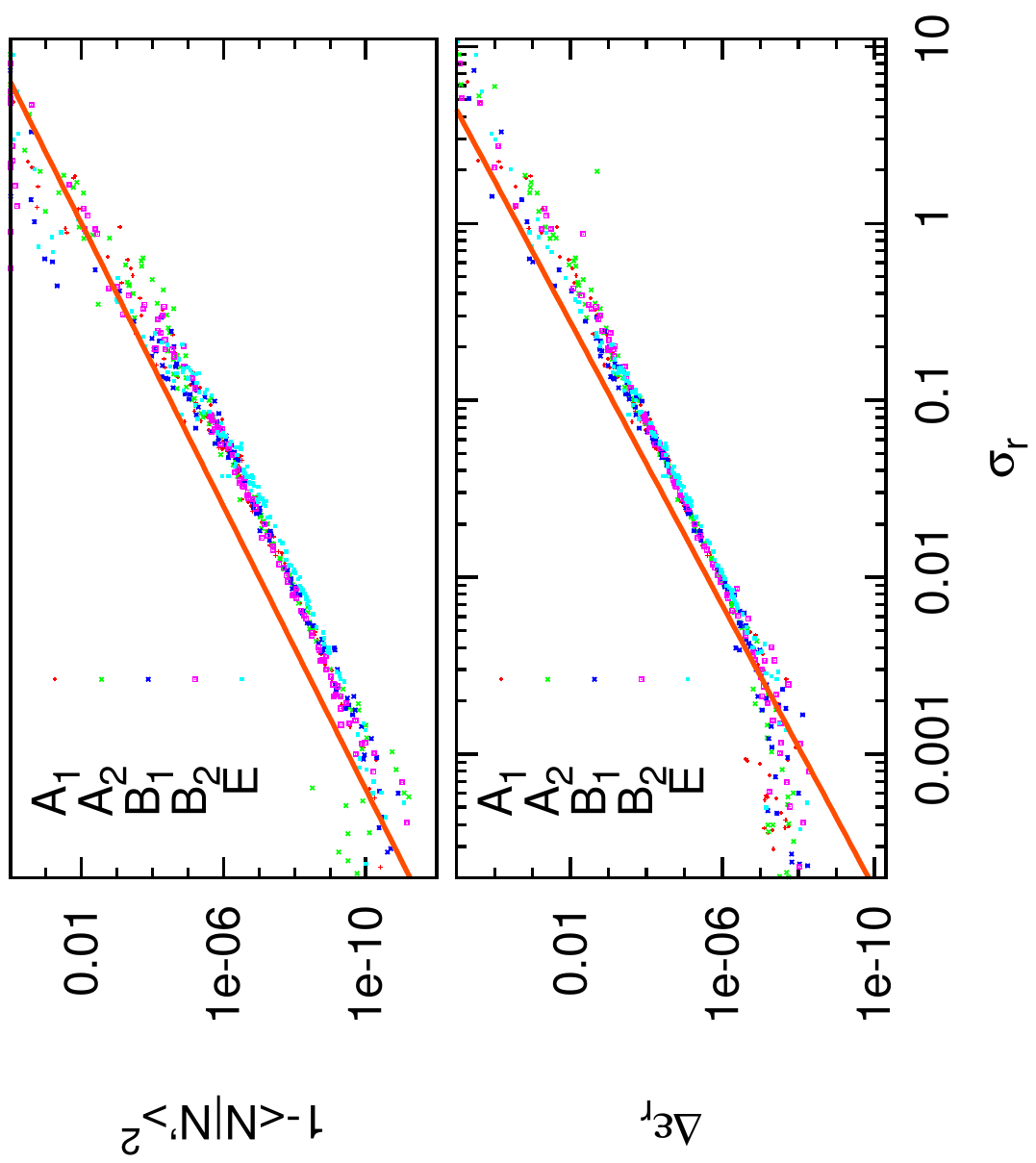}
\caption{(Color online)
Error in the eigenenergies (bottom panel) and eigenfunctions (top panel)
of the eigenstates obtained in our calculation using a basis set of scar functions,
estimated as described in subsection \ref{subsec:error},
as a function of the relative dispersion, $\sigma_r=\sigma_N \rho$.
The continuous lines indicate the upper error bounds given in Eq.~(\ref{eq:47}).}
\label{fig:20}
\end{figure}

In Fig.~\ref{fig:20} we present such errors for both
eigenenergies (bottom panel) and eigenfunctions (top panel)
of all symmetry classes,
computed as $\Delta \E_r=\vert \E_N - \E_{N'} \vert \rho$
(measured in mean energy level spacing units of each symmetry-class)
and $1-\langle N \vert N' \rangle^2$, respectively,
as a function of the reduced dispersion of the eigenfunctions
$$\sigma_r=\sigma_N \rho,$$ 
where $\sigma_N$ is the dispersion of the eigenfunction $\vert N \rangle_\chi$.
As can be seen, both 
$\Delta \E_r$ and $1-\langle N \vert N' \rangle^2$
follow to a great degree of accuracy
a power law for $\sigma_r<0.3$, thus giving upper bounds that can be
expressed as
%
\begin{equation}
\Delta \E_r \le \frac{\sigma_r^{2.5}}{4}, \qquad
1-\langle N \vert N' \rangle^2 \le \frac{\sigma_r^{2.5}}{100}.
\label{eq:47}
\end{equation}
There is a good reason for calculating the previous upper bounds 
as a function of  $\sigma_r$.
Indeed, this latter magnitude can be computed straightforwardly, 
so that one can estimate the errors in the results with the aid 
of Eq.~(\ref{eq:47}), without the need of any further calculation.
Notice that for very small values of the dispersion,
the accuracy of our calculations is dominated by precision errors,
and these expressions for the error bounds do not apply.

%
%
\section{Conclusions}
\label{sec:summary}

In this paper, we develop a new method to construct basis sets from
scar functions which are able to calculate very accurately the
eigenstates of classically chaotic systems and keeping the size
of the problem at very moderate sizes.
The main idea is that introducing the relevant dynamical information
into the basis elements has a twofold effect.
First, the efficiency of the basis set is fostered, and second,
its nature allows us an straightforward identification of the underlying
classical structures contributing to each individual eigenstate,
thus providing a good description of the quantum dynamics in a
semiclassical sense.

As an illustration we applied the method to a classically chaotic
quartic two-dimensional oscillator [see Eq.~(\ref{eq:1})],
that has been used as a benchmark in the field of quantum chaos.
The performance of our method for this system was superb,
since we have demonstrated that the method can be advantageously 
applied to the computation of excited states in a small energy window 
using very modest basis sets, whose sizes can be estimated from
the average participation ratio. Moreover, 
using a basis set of $\sim$2500 elements,
we have been able to compute the first low-lying $\sim$2400 eigenstates 
with an accuracy, both in energies and wave functions, similar to that
of other standard methods used in the past to study the same oscillator.

Furthermore, we have examined the quality of our results using a variety
of different indicators, such as projection on local (scar) representations,
scar intensities, or participation ratios.
The first two allow a direct description of the different eigenstates
in terms of semiclassical ideas, while the latter provides an idea
of how our scar basis is much smaller than other conventional one.
Furthermore, we also provided in this work upper bounds to the errors
that can be expected in our calculations, both in energy and wave functions.

We are currently extending the method to realistic molecular
systems with a mixed phase space, where the degree of chaoticity
depends on the energy.
The results of an application to the LiNC/LiCN isomerizing system
will be reported elsewhere.

\section{ACKNOWLEDGEMENTS}
\label{sec:acknow}
This work has been supported by MINECO (Spain)
under projects MTM2009-14621 and ICMAT Severo
Ochoa SEV-2011-0087,
and CEAL Banco de Santander--UAM.
We thank the referees, whose comments 
helped to improve the presentation of the paper.
FR gratefully thanks a doctoral fellowship form 
UPM and the hospitality
of the members of the Departamento de F\'isica 
in the Laboratorio TANDAR--Comisi\'on 
Nacional de la Energ\'ia At\'omica,
where part of this work was done.


\newpage
\appendix
\newpage
\section*{Supplemental material}

\begin{figure*}
\includegraphics{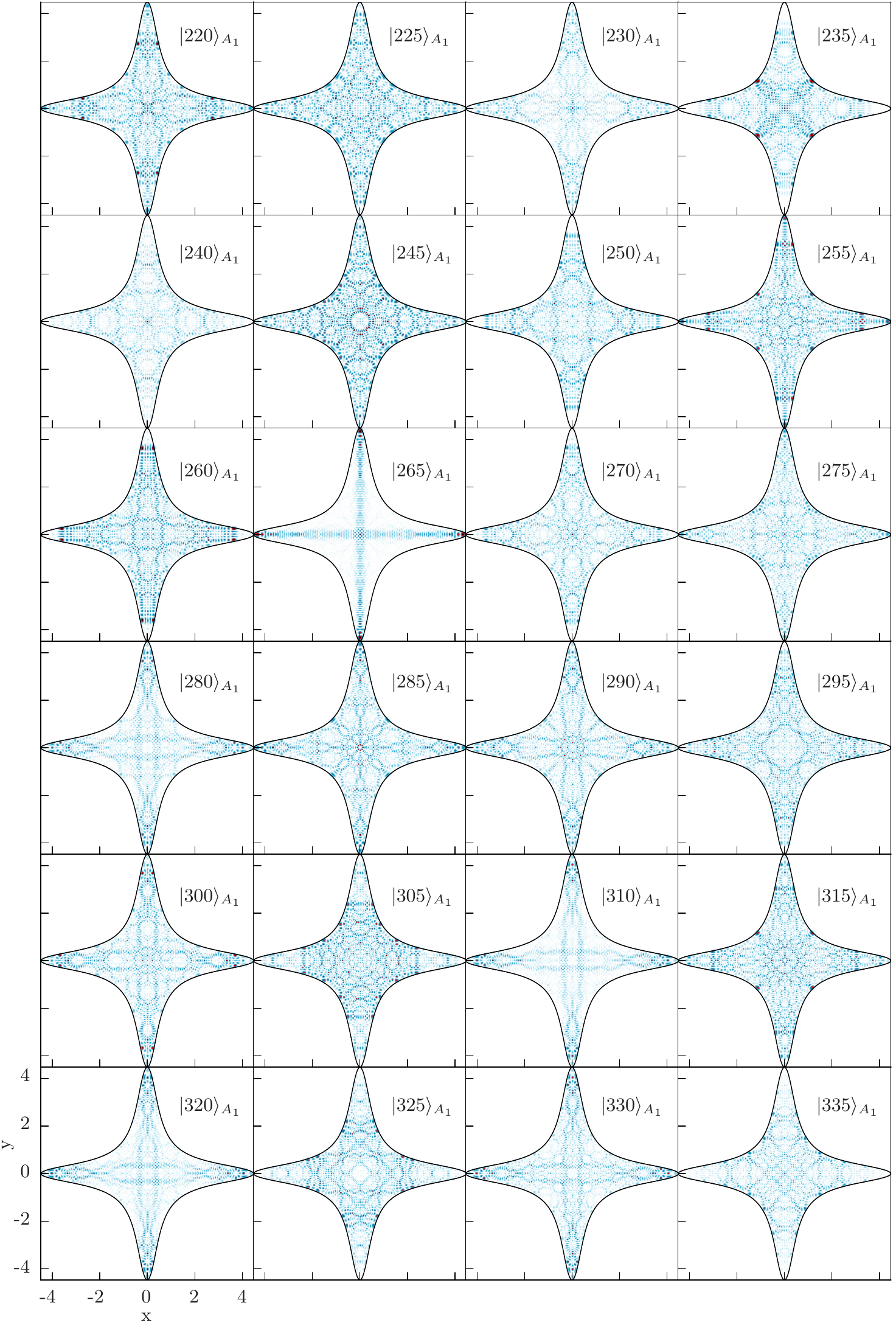}
\caption{(Color online) Quartic oscillator eigenfunctions $\vert N \rangle_{A_1}$
corresponding to the reconstruction data of Table~\ref{Table:IV}.
Same coordinates and scaling as in Fig.~5.}
\label{fig:13}
\end{figure*}

%
%
Here, we present some additional material containing our results on the structure of
the eigenfunctions of the quartic oscillator in a more systematic way.
The relevant numerical data corresponding to the reconstruction
with scar functions of the eigenfunctions~$\vert 220+5i \rangle_{A_1}$,
with~$i=0,1,\ldots,23$, whose probability densities are shown in
Fig.~\ref{fig:13}, are reported in Table~\ref{Table:IV}.
In it, we have included all scar functions necessary to reconstruct, in each case,
more than $85\%$ of the corresponding eigenfunctions,
remark that all computed states were obtained with the same accuracy as obtained
by other standard methods~\cite{pullenedmonds81_carne84_EHP89,Waterland}.
As we can see in the figure, the pattern of several eigenfunctions is
concentrated only over a single PO of the system.
In these cases, the first element of the \emph{local} representation coincides
with the scar function that is localized over that specific PO,
similarly to what happens with the eigenfunctions in
Figs.~9--11.
In the rest of the cases, and although the eigenfunctions are in general
not scarred by a single PO, the SGSM does still permit their reconstruction
using a few (localized) scar functions of our basis set,
as can be checked in the data presented in Table~\ref{Table:IV}.

\begin{figure*}
\includegraphics{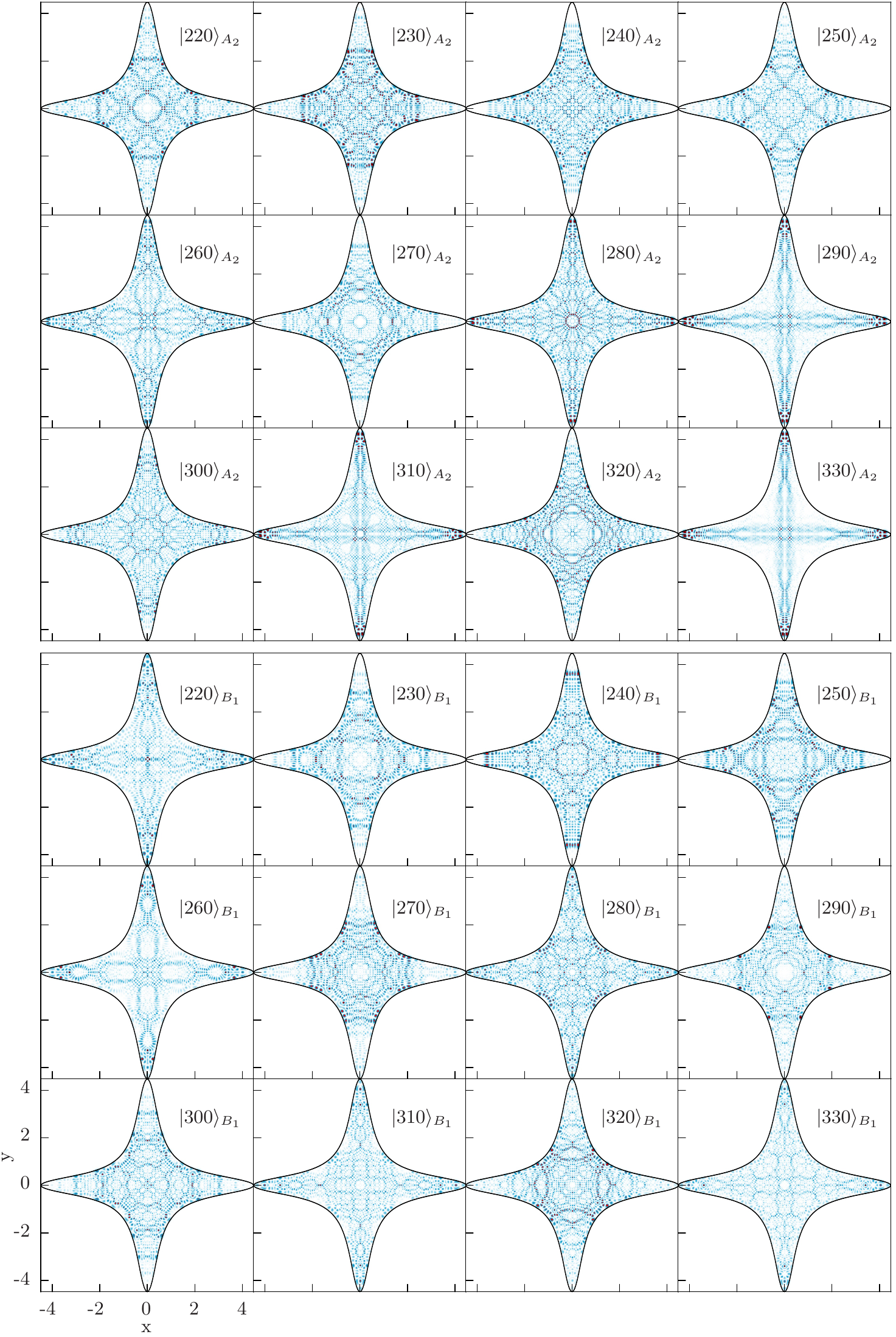}
\caption{(Color online)
Same as Fig.~\ref{fig:13} for the eigenfunctions $\vert N \rangle_{A_2}$ and $\vert N \rangle_{B_1}$
corresponding to the reconstruction data of Tables~\ref{Table:V} and \ref{Table:VI}.}
\label{fig:14}
\end{figure*}

\begin{figure*}
\includegraphics{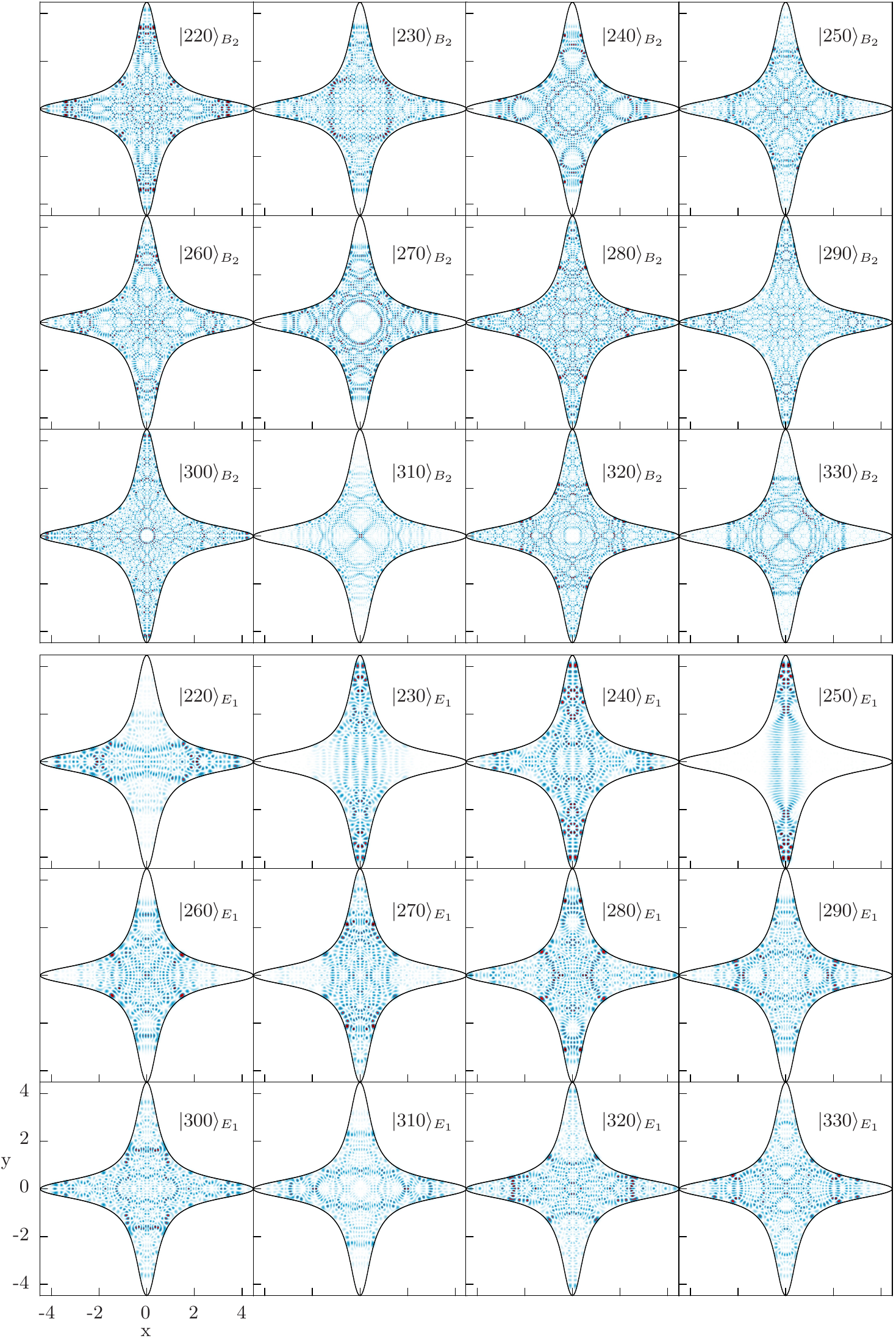}
\caption{(Color online)
Same as Fig.~\ref{fig:13} for the eigenfunctions $\vert N \rangle_{B_2}$ and $\vert N \rangle_{E_1}$
corresponding to the reconstruction data of Tables~\ref{Table:VII} and \ref{Table:VIII}.
}
\label{fig:15}
\end{figure*}

In a similar way, we present in Tables~\ref{Table:V}--\ref{Table:VIII}
the results corresponding to the eigenfunctions in 
Figs.~\ref{fig:14}--\ref{fig:15},
associated to the other symmetry classes.

\begin{table*}
\caption{Reconstruction of the eigenstates $\vert N \rangle_{A_1}$ with the scar functions
$\vert {\rm PO}, n \rangle_{A_1}$.
The participation ratio, $\PR$, defined in Eq.~(35) and the accumulated reconstruction overlap, $\Sigma\%$,
are also given.}
%
\begin{tabular}{rrrrrrrrrrrr}
\hline\hline
$N$ & $\PR$ & PO,$n$,$\Sigma\%$ & & & & & & & & & \\
\hline
220 & 2.37 & 14,44,62.5 & 4,57,77.6 & 2,48,82.9 & 9,47,86.5 & & & & & & \\
225 & 3.83 & 4,58,35.3 & 9,48,63.0 & 6,28,86.7 & & & & & & & \\
230 & 2.71 & 14,45,57.6 & 2,49,74.0 & 7,21,77.7 & 6,28,81.3 & 16,38,85.6 & & & & & \\
235 & 6.45 & 3,19,20.0 & 4,59,43.5 & 17,36,61.3 & 7,21,76.3 & 14,46,79.1 & 14,45,81.3 & 16,39,82.4 & 13,33,83.4 & 5,12,84.4 & 6,28,85.1 \\
240 & 3.45 & 14,46,49.8 & 5,12,62.6 & 16,39,75.9 & 13,33,82.4 & 4,59,86.1 & & & & & \\
245 & 6.43 & 12,30,29.5 & 4,61,42.5 & 17,37,52.9 & 6,29,58.4 & 16,40,63.4 & 1,51,66.8 & 2,51,82.0 & 9,50,86.1 & & \\
250 & 3.83 & 16,40,46.7 & 12,30,63.3 & 9,51,70.5 & 17,37,75.2 & 7,22,78.8 & 2,51,82.3 & 1,51,85.6 & & & \\
255 & 6.52 & 1,52,26.1 & 4,62,43.9 & 14,48,61.0 & 3,20,72.3 & 13,34,78.9 & 7,22,85.7 & & & & \\
260 & 3.56 & 9,52,49.7 & 14,48,58.3 & 15,46,68.4 & 6,30,76.6 & 4,63,82.8 & 12,31,88.1 & & & & \\
265 & 1.30 & 1,53,87.7 & & & & & & & & & \\
270 & 4.60 & 9,53,36.6 & 4,64,59.2 & 17,39,72.1 & 8,31,80.5 & 13,35,87.4 & & & & & \\
275 & 5.25 & 2,54,31.0 & 5,13,55.9 & 8,31,69.0 & 13,35,76.0 & 4,64,84.9 & 17,39,87.5 & & & & \\
280 & 2.15 & 9,54,67.7 & 11,52,70.9 & 17,40,72.9 & 13,36,75.5 & 10,53,76.4 & 5,13,77.2 & 10,54,77.9 & 9,55,79.4 & 4,66,80.9 & 1,55,81.9 \\
& & 2,55,86.4 & & & & & & & & & \\
285 & 4.27 & 2,55,41.0 & 13,36,59.9 & 3,21,71.6 & 17,40,78.6 & 4,65,88.7 & & & & & \\
290 & 6.87 & 9,55,29.3 & 4,66,39.6 & 10,54,47.0 & 15,49,49.5 & 6,32,51.3 & 8,32,67.2 & 7,24,69.7 & 17,41,71.5 & 4,67,72.9 & 11,52,73.8 \\
& & 13,36,74.7 & 16,43,75.9 & 1,56,76.6 & 2,56,86.7 & & & & & & \\
295 & 3.45 & 4,67,48.9 & 10,55,60.9 & 9,56,75.6 & 17,41,85.1 & & & & & & \\
300 & 2.50 & 9,56,58.9 & 16,44,73.9 & 10,55,90.8 & & & & & & & \\
305 & 8.03 & 4,68,13.4 & 18,48,26.9 & 10,56,42.3 & 11,54,56.6 & 9,57,70.4 & 3,22,85.6 & & & & \\
310 & 2.44 & 9,57,62.8 & 11,54,69.5 & 6,33,73.2 & 18,48,76.0 & 10,56,77.4 & 1,58,78.7 & 2,58,86.1 & & & \\
315 & 8.96 & 11,55,23.0 & 2,58,41.3 & 12,34,49.6 & 4,69,54.3 & 5,14,58.1 & 1,58,60.3 & 9,57,67.8 & 11,54,69.7 & 7,25,71.2 & 18,48,72.6 \\
& & 16,45,74.2 & 6,33,75.5 & 3,22,76.3 & 10,57,77.0 & 9,58,80.3 & 18,49,86.9 & & & & \\
320 & 3.60 & 9,58,49.0 & 11,55,57.3 & 4,70,63.2 & 5,14,65.1 & 1,59,66.7 & 2,59,81.9 & 11,56,84.1 & 17,43,87.0 & & \\
325 & 2.93 & 16,46,52.1 & 8,34,77.4 & 17,43,82.8 & 7,25,86.3 & & & & & & \\
330 & 9.38 & 10,58,18.9 & 4,71,33.0 & 18,50,39.7 & 11,56,44.2 & 6,34,48.2 & 16,46,56.5 & 8,34,61.3 & 1,60,64.5 & 2,60,81.2 & 13,39,82.3 \\
& & 3,23,84.5 & 16,47,87.1 & & & & & & & & \\
335 & 6.55 & 18,50,35.1 & 17,44,42.8 & 11,57,50.1 & 4,71,54.4 & 2,60,58.4 & 10,58,62.8 & 1,60,67.0 & 11,56,70.7 & 16,46,74.4 & 6,34,80.1 \\
& & 13,39,84.6 & 2,59,87.4 & & & & & & & \\
\hline
\end{tabular}
\label{Table:IV}
\end{table*}
%
%
%
\begin{table*}
\caption{Same as Table~\ref{Table:IV} for the eigenfunctions $\vert N \rangle_{A_2}$.}
\begin{tabular}{rrrrrrrrrrrr}
\hline\hline
$N$ & $\PR$ & PO,$n$,$\Sigma\%$ & & & & & & & & & \\
\hline
220 & 4.67 & 6,28,40.5 & 4,58,49.7 & 7,21,62.9 & 9,47,68.5 & 2,48,79.7 & 11,45,88.1 & & & & \\
230 & 3.17 & 4,59,43.3 & 18,42,77.9 & 10,48,79.4 & 10,49,81.0 & 11,46,82.4 & 9,48,90.6 & & & & \\
240 & 4.62 & 14,47,43.8 & 10,50,53.1 & 4,60,60.2 & 5,12,64.0 & 6,29,66.8 & 17,38,68.9 & 7,22,72.9 & 12,31,76.2 & 16,40,79.8 & 2,51,83.3 \\
& & 8,30,85.1 & & & & & & & & & \\
250 & 4.76 & 8,30,37.7 & 4,62,56.9 & 17,39,65.3 & 10,51,73.8 & 6,30,84.4 & 16,41,88.8 & & & & \\
260 & 4.19 & 14,49,39.7 & 2,53,53.3 & 10,52,76.3 & 12,32,82.2 & 9,52,85.8 & & & & & \\
270 & 4.13 & 16,42,42.7 & 13,35,49.8 & 17,40,59.9 & 14,50,79.4 & 4,64,85.2 & & & & & \\
280 & 3.79 & 2,55,38.5 & 12,33,68.9 & 14,51,81.5 & 4,66,85.2 & & & & & & \\
290 & 1.49 & 2,56,81.5 & 4,67,86.1 & & & & & & & & \\
300 & 3.56 & 14,53,47.7 & 10,56,65.1 & 8,33,77.8 & 18,49,82.1 & 9,56,85.1 & & & & & \\
310 & 1.75 & 2,58,72.2 & 13,38,94.4 & & & & & & & & \\
320 & 4.90 & 6,34,33.0 & 18,50,62.0 & 11,55,66.7 & 2,59,70.9 & 15,52,74.7 & 9,57,77.5 & 5,14,81.6 & 2,58,84.6 & 4,70,87.2& \\
330 & 1.22 & 2,60,90.5 & & & & & & & & & \\
\hline
\end{tabular}
\label{Table:V}
\end{table*}
%
\begin{table*}
\caption{Same as Table~\ref{Table:IV} for eigenfunctions $\vert N \rangle_{B_1}$.}
\begin{tabular}{rrrrrrrrrrrrr}
\hline\hline
$N$ & $\PR$ & PO,$n$,$\Sigma\%$ & & & & & & & & & \\
\hline
220 & 3.03 & 14,45,46.8 & 2,49,79.4 & 16,37,84.3 & 17,36,87.7 & & & & & & \\
230 & 5.20 & 4,59,31.4 & 16,38,58.6 & 17,37,66.7 & 7,21,73.0 & 6,28,76.8 & 2,50,79.1 & 1,50,82.2 & 9,48,89.3 & & \\
240 & 7.76 & 6,29,29.2 & 9,50,45.3 & 16,39,54.2 & 5,12,57.6 & 11,48,62.1 & 2,51,65.9 & 14,47,68.4 & 12,30,71.3 & 10,51,72.8 & 9,51,74.6 \\
& & 13,33,76.3 & 17,38,77.9 & 4,61,79.3 & 7,22,81.8 & 4,60,83.2 & 2,52,84.6 & 12,31,85.8 & & & \\
250 & 2.36 & 16,40,63.9 & 11,49,71.3 & 7,22,77.9 & 13,33,79.0 & 17,38,83.9 & 5,12,86.7 & & & & \\
260 & 1.76 & 9,52,73.5 & 17,39,89.7 & & & & & & & & \\
270 & 2.35 & 8,31,64.2 & 18,46,70.1 & 12,32,76.7 & 4,64,81.2 & 9,53,83.7 & 14,51,85.2 & & & & \\
280 & 6.48 & 5,13,25.9 & 1,56,36.8 & 2,56,54.5 & 9,54,72.1 & 12,33,78.7 & 18,47,87.3 & & & & \\
290 & 4.40 & 6,32,43.0 & 13,36,58.3 & 4,67,65.9 & 5,13,72.9 & 11,52,74.5 & 4,66,76.2 & 11,53,77.4 & 1,57,78.6 & 2,57,83.8 & 9,55,88.4 \\
300 & 2.25 & 13,37,63.3 & 4,68,83.7 & 18,48,87.6 & & & & & & & \\
310 & 8.11 & 10,57,14.0 & 1,59,19.1 & 2,59,47.7 & 9,57,54.8 & 11,55,61.2 & 11,54,65.0 & 4,69,68.3 & 17,43,72.2 & 13,37,74.7 & 18,49,77.2 \\
& & 6,33,80.5 & 12,35,83.5 & 2,58,86.5 & & & & & & & \\
320 & 7.56 & 12,35,28.3 & 4,70,42.7 & 13,38,53.9 & 18,50,59.9 & 8,34,65.4 & 5,14,71.4 & 6,34,75.8 & 1,60,78.0 & 2,60,81.9 & 9,58,86.8 \\
330 & 7.09 & 10,59,25.5 & 4,71,35.7 & 17,44,41.2 & 1,61,46.0 & 2,61,69.4 & 5,14,73.8 & 8,34,75.5 & 6,34,79.4 & 12,35,81.1 & 13,38,82.2 \\
& & 4,70,83.5 & 17,43,84.3 & 11,57,85.0 & & & & & & & \\
\hline
\end{tabular}
\label{Table:VI}
\end{table*}
%
%
\begin{table*}
\caption{Same as Table~\ref{Table:IV} for eigenfunctions $\vert N \rangle_{B_2}$.}
\begin{tabular}{rrrrrrrrrrrr}
\hline\hline
$N$ & $\PR$ & PO,$n$,$\Sigma\%$ & & & & & & & & & \\
\hline
220 & 7.55 & 10,46,27.1 & 4,57,32.8 & 9,47,36.9 & 2,47,51.7 & 11,45,67.2 & 13,32,70.0 & 10,47,72.1 & 6,28,73.5 & 8,28,79.0 & 3,19,80.8 \\
& & 2,48,82.1 & 5,12,83.4 & 11,46,84.1 & 9,48,88.9 & & & & & & \\
230 & 4.30 & 18,41,37.4 & 5,12,64.4 & 10,47,75.7 & 8,28,76.8 & 9,48,77.8 & 11,46,84.1 & 2,48,87.6 & & & \\
240 & 1.53 & 4,60,79.7 & 18,42,92.8 & & & & & & & & \\
250 & 2.72 & 18,43,50.0 & 14,47,83.6 & 16,40,88.5 & & & & & & & \\
260 & 2.00 & 14,48,70.3 & 11,50,73.5 & 9,51,75.0 & 5,13,76.4 & 10,50,77.5 & 2,51,80.5 & 10,51,82.3 & 14,47,83.4 & 4,62,84.4 & 11,49,86.5 \\
270 & 3.26 & 4,64,47.4 & 6,31,71.7 & 14,49,82.5 & 16,42,92.9 & & & & & & \\
280 & 3.93 & 16,43,44.0 & 17,40,61.8 & 10,53,76.1 & 2,54,82.9 & 14,50,84.9 & 9,54,89.2 & & & & \\
290 & 3.79 & 14,51,46.7 & 17,41,60.7 & 4,67,67.6 & 15,49,75.0 & 7,24,86.2 & & & & & \\
300 & 8.66 & 14,52,22.3 & 2,56,39.4 & 6,33,47.1 & 10,55,53.5 & 9,56,62.8 & 4,68,68.9 & 3,22,76.1 & 18,48,78.8 & 15,50,83.4 & 5,14,90.3 \\
310 & 3.71 & 12,34,45.0 & 4,69,67.3 & 7,25,76.5 & 15,51,83.0 & 6,33,87.6 & & & & & \\
320 & 4.58 & 13,39,32.9 & 6,34,61.3 & 10,57,69.2 & 2,58,74.8 & 12,35,79.9 & 14,54,82.9 & 9,58,95.7 & & & \\
330 & 4.02 & 16,47,46.2 & 3,23,55.2 & 17,44,63.6 & 10,58,72.2 & 7,26,77.8 & 18,50,84.4 & 13,40,90.5 & & & \\
\hline
\end{tabular}
\label{Table:VII}
\end{table*}

\begin{table*}
\caption{Same as Table~\ref{Table:IV} for eigenstates $\vert N \rangle_{E_1}$.}
\begin{tabular}{rrrrrrrrrrrr}
\hline\hline
$N$ & $\PR$ & PO,$n$,$\Sigma\%$ & & & & & & & & & \\
\hline
220 & 4.47 & 9,34,40.2 & 7,27,54.3 & 2,31,67.6 & 14,32,81.8 & 8,39,86.2 & & & & & \\
230 & 4.53 & $\widetilde{10}$,32,42.9 & 13,23,55.6 & $\widetilde{12}$,20,66.0 & $\widetilde{2}$,35,73.7 & $\widetilde{4}$,40,77.4 & 8,39,79.0 & $\widetilde{15}$,29,79.8 & 6,40,80.7 & 4,42,82.4 & 11,33,83.8 \\
& & 9,35,84.8 & 9,34,85.8 & & & & & & & & \\
240 & 4.34 & $\widetilde{15}$,30,30.3 & $\widetilde{2}$,36,58.4 & 11,34,81.7 & 5,17,85.7 & & & & & & \\
250 & 2.15 & $\widetilde{2}$,37,62.3 & $\widetilde{15}$,31,89.5 & & & & & & & & \\
260 & 2.85 & $\widetilde{4}$,43,52.8 & 12,22,78.3 & 5,18,83.6 & 4,44,87.4 & & & & & & \\
270 & 4.35 & $\widetilde{12}$,22,40.3 & $\widetilde{4}$,44,62.4 & 6,43,70.8 & 13,25,76.3 & $\widetilde{10}$,35,81.8 & 8,44,86.8 & & & & \\
280 & 6.75 & $\widetilde{4}$,45,31.4 & 7,31,46.4 & 2,36,56.1 & 12,23,64.4 & $\widetilde{14}$,34,69.9 & $\widetilde{10}$,36,76.8 & 9,39,80.1 & 10,38,31.4 & 3,15,82.5 & $\widetilde{2}$,40,83.3 \\
& & 9,38,83.8 & 15,34,84.7 & $\widetilde{11}$,38,85.2 & & & & & & & \\
290 & 6.35 & 15,35,28.3 & 6,45,43.1 & $\widetilde{4}$,46,61.4 & $\widetilde{12}$,23,73.2 & 3,15,76.2 & 13,26,83.2 & $\widetilde{11}$,38,85.9 & & & \\
300 & 6.94 & 9,40,27.2 & 7,32,45.4 & $\widetilde{12}$,23,60.2 & $\widetilde{4}$,46,66.0 & 15,35,69.3 & 2,37,73.0 & 4,48,75.6 & $\widetilde{11}$,38,77.8 & $\widetilde{10}$,37,85.0 & \\
310 & 4.11 & 15,36,47.3 & 12,24,55.3 & $\widetilde{11}$,39,60.2 & 6,46,63.0 & 8,46,71.3 & 10,40,43.5 & 9,40,75.2 & $\widetilde{14}$,36,76.9 & $\widetilde{9}$,38,79.0 & $\widetilde{12}$,23,80.8 \\
& & 4,48,82.1 & 7,33,83.1 & 15,35,84.1 & 7,32,84.5 & 13,27,84.8 & 4,49,85.2 & & & & \\
320 & 6.70 & 4,50,28.0 & 3,16,41.1 & 10,41,55.4 & 15,37,66.1 & 2,39,76.9 & $\widetilde{2}$,42,85.7 & & & & \\
330 & 9.24 & $\widetilde{4}$,49,19.3 & 12,25,29.6 & 4,50,37.7 & 6,20,46.2 & 13,28,63.1 & 3,16,75.1 & 4,51,78.3 & $\widetilde{9}$,40,79.7 & $\widetilde{2}$,43,82.0 & $\widetilde{14}$,37,83.2 \\
& & $\widetilde{11}$,40,85.0 & & & & & & & & & \\
\hline
\end{tabular}
\label{Table:VIII}
\end{table*}
%

\end{document}